\documentclass[a4paper,preprintnumbers,showpacs,onecolumn,superscriptaddress,nofootinbib,amsmath,amssymb,notitlepage]{revtex4-1}

\usepackage{enumitem}
\usepackage{times}
\usepackage{dcolumn}
\usepackage{mathrsfs}

\usepackage{comment}
\usepackage{hyperref}
\usepackage{amsmath}
\usepackage{mathtools}
\usepackage{bbm}
\usepackage{bm}
\usepackage{amsfonts}
\usepackage{amssymb}
\usepackage{mathrsfs}
\usepackage{wasysym}
\usepackage{graphicx}
\usepackage[]{subfigure}
\usepackage{filecontents}
\usepackage[dvipsnames]{xcolor}

\hypersetup{
    bookmarks=true,         
    pdftoolbar=true,        
    pdfmenubar=true,        
    pdffitwindow=true,     
    pdfstartview={FitH},     
    pdftitle={My title},     
    pdfauthor={author},      
    pdfsubject={Subject},    
    pdfcreator={Creator},    
    pdfproducer={Producer},  
    pdfkeywords={keyword1} 
    pdfnewwindow=true,       
    colorlinks=true ,        
    linkcolor=Blue  ,        
    citecolor=Blue,      
    filecolor=Blue,       
    urlcolor=Blue            
}

\newcommand{\arccosh}{\operatorname{arccosh}}

\newcommand{\ed}{\mathrm{d}}

\newcommand{\Q}{\mathscr{Q}}

\begin{document}

\title{Spherical particle orbits around a rotating black hole in massive gravity
}

\author{Mohsen Fathi}
\email{mohsen.fathi@gmail.com; mohsen.fathi@usach.cl}
\affiliation{Departamento de F\'{i}sica, Universidad de Santiago de Chile,
Avenida V\'{i}ctor Jara 3493,  Estaci\'{o}n Central, 9170124, Santiago, Chile}

\author{J.R. Villanueva}
\email{jose.villanueva@uv.cl}
\affiliation{Instituto de F\'{i}sica y Astronom\'{i}a, Universidad de Valpara\'{i}so,
Avenida Gran Breta\~{n}a 1111, Valpara\'{i}so, Chile}

\author{Norman Cruz}
\email{norman.cruz@usach.cl}
\affiliation{Departamento de F\'{i}sica, Universidad de Santiago de Chile,
Avenida V\'{i}ctor Jara 3493,  Estaci\'{o}n Central, 9170124, Santiago, Chile}

\begin{abstract}

In this paper, we present a rotating de Rham-Gabadadze-Tolley black hole with a positive cosmological constant in massive gravity, achieved by applying a modified Newman-Janis algorithm. The black hole exhibits stable orbits of constant radii, prompting a numerical study on the behavior of the solutions to a nonic equation governing the radii of planar orbits around the black hole. Additionally, we investigate the stability of orbits near the event horizon and provide a comprehensive analytical examination of the solutions to the angular equations of motion. This is followed by simulating some spherical particle orbits around the black hole.


\bigskip

{\noindent{\textit{keywords}}: Particle geodesics, massive gravity, cosmological constant, ISCO 
}\\

\end{abstract}

\maketitle

\tableofcontents

\section{Introduction and Motivation}\label{sec:intro}

The significance of investigating stationary black hole spacetimes derived from general relativity is undeniable. Among these, the Kerr-like black holes have garnered significant attention, particularly following the groundbreaking detections of gravitational waves from a merger by the Laser Interferometer Gravitational-Wave Observatory (LIGO) \cite{PhysRevLett.116.061102} and the remarkable imaging of black holes M87* and Sgr A* by the Event Horizon Telescope (EHT) \cite{EventHorizonTelescope:2019ths,EventHorizonTelescope:2022xnr}. Subsequently, there has been a surge in the study of geodesics involving massless and massive particles in the vicinity of Kerr-like black holes, as evidenced by recent publications (see for example, Refs. \cite{rana_astrophysically_2019,kapec_particle_2020,gralla_null_2020,stein_location_2020,compere_near-horizon_2020,rana_geometric_2020}). However, the study of particle motion around Kerr black holes dates back to 1968 with the introduction of Carter's method for separating the Hamilton-Jacobi equation \cite{Carter:1968}. This method allows orbits to be classified as either planar or non-planar based on variations in their polar component. While the resulting first-order differential equations of motion can be easily solved numerically, finding exact analytical solutions for the coordinate evolution is challenging due to their nonlinear nature. Extensive research has been conducted to address this issue. For example, Mino demonstrated that by introducing a new time parameter, two of the four equations of motion can be decoupled \cite{Mino:2003}, leading to solutions expressed in terms of elliptic integrals \cite{fujita_analytical_2009} (also discussed in the review article \cite{nicolini_analytical_2016} and references therein). Moreover, when the radial coordinate is held constant, the resulting orbits are either circular on the equatorial plane or completely non-planar, forming spherical orbits. In fact, the study of such orbits is valuable in astrophysical research as they help determine the specific domains where light and particles are captured by the black hole from different polar angles. Numerous studies have focused on determining spherical particle orbits characterized by time-like constant-radius geodesics, which follow the pioneering work by Wilkins \cite{PhysRevD.5.814}. Since then, extensive investigations have explored various aspects of these orbits that provide insights into their properties and behavior (see for example, Refs. \cite{rana_astrophysically_2019,stein_location_2020,compere_near-horizon_2020,PhysRevD.10.2324,stoghianidis_polar_1987,hughes_evolution_2000,hughes_evolution_2001,kraniotis_precise_2004,fayos_geometrical_2008,hackmann_analytical_2010,grossman_harmonic_2012,hod_marginally_2013,teo_spherical_2021,Tavlayan:2021,battista_geodesic_2022}). In these studies, one often encounters algebraically complex equations, such as the polynomial equations that govern the radii of spherical orbits. Consequently, both rigorous analytical and numerical approaches are necessary to determine reliable ranges for the motion parameters that satisfy the orbital conditions in general relativistic spacetimes, which are crucial for unveiling accurate insights into the behavior and constraints of spherical orbits in the context of general relativity.

However, while general relativity and its black hole solutions have been successful in numerous observational tests, there are still unanswered questions, particularly regarding the mysterious dark side of the universe \cite{RevModPhys.75.1433}, which remains one of the major enigmas in modern cosmology. In response, some scientists suggest that extending general relativity in the right way could provide an alternative approach that may eliminate the need for dark matter and dark energy, offering new perspectives and potential solutions to these intriguing phenomena. Such extensions include the $f(R)$ \cite{RevModPhys.82.451,de_felice_fr_2010} and scalar-tensor theories of gravity \cite{quiros_selected_2019,fujii_maeda_2003},
in the four-dimensional spacetimes. In particular, massive gravity has received significant attention, tracing back to the Lorentz invariant Fierz-Pauli massive spin-2 theory in 1939 \cite{doi:10.1098/rspa.1939.0140}. Since then, this theory has been further developed and generalized by ed Rham, Gabadadze, and Tolley (dRGT) \cite{PhysRevD.82.044020,PhysRevLett.106.231101} (see also the review \cite{de_rham_massive_2014}). In Ref. \cite{ghosh_class_2016}, a spherically symmetric black hole solution within the dRGT theory was presented. By generalizing the Schwarzschild spacetime, this latter solution incorporates the new parameter $\gamma$ that can potentially account for the flat galactic rotation curves, in the sense that the massive gravitons constitute a dark matter halo \cite{Panpanich:2018}. Furthermore, the aforementioned solution is endowed with a positive cosmological constant, which serves to compensate for the accelerated expansion of the universe. In fact, the presence of a positive cosmological constant has been argued to have significant gravitational effects, not only on black hole dynamics but also on the cosmos as a whole \cite{Ashtekar_2017}. In a related study, it was demonstrated in Ref. \cite{Visser:2020} that the inclusion of a positive cosmological constant enables black hole spacetimes to exhibit innermost and outermost stable circular orbits (ISCOs and OSCOs), which have important implications for the behavior and stability of particles around black holes.

In line with the same research interest, our paper investigates a black hole within the framework of dRGT massive gravity, specifically considering the form proposed in Ref.~\cite{Panpanich:2018}. This black hole solution incorporates a positive cosmological constant, allowing for the formation of an ISCO and an OSCO, as recently studied in Ref.~\cite{universe7080278}. However, as highlighted in the opening of this section, it is worth noting that stable circular orbits represent only a subset of the broader category of spherical particle orbits, which forms the primary focus of our investigation. In this regard, we construct a rotating counterpart of the aforementioned dRGT spacetime and explore various facets of spherical time-like geodesics within the black hole's exterior region. Consequently, this paper centers on two key objectives; examining the radii of spherical orbits and their corresponding orbit stability, and presenting analytical solutions to the angular equations of motion for a comprehensive analysis. To achieve these objectives, the paper is organized as follows: In Sect. \ref{sec:BlackHole}, we provide a concise introduction to the dRGT massive gravity theory and its static black hole solution. Subsequently, we present the rotating counterpart of this black hole spacetime and analyze its causal structure. Moving on to Sect. \ref{sec:geodesic}, we initiate the investigation of spherical orbits using Carter's method of separation of the Hamilton-Jacobi equation. By applying the general criteria for the formation of spherical orbits, we obtain a polynomial equation of order fourteen. However, for the purpose of this investigation, we focus specifically on planar orbits and numerically analyze the solutions to the corresponding characteristic equation. In Section \ref{sec:AnlyticalSol}, we tackle the problem by solving the first-order nonlinear differential equations governing the polar and azimuth angles. This allows us to derive exact analytical solutions for the constant-radius equations of motion that describe spherical orbits. These solutions are expressed in terms of Weierstrassian elliptic functions, which are then employed to showcase illustrative examples of orbits around the black hole. Finally, in Sect. \ref{sec:conclusion}, we summarize our findings. Throughout this study, we adopt geometrized units where $G=c=1$, employ the sign convention $(- + +\, +)$ for the spacetime line element, and denote differentiations with respect to $r$ or $x$-coordinates using primes.

\section{Massive theory of gravity and its static black hole solution in the cosmological background}\label{sec:BlackHole}

Considering the Riemannian manifold $(\mathcal{M},g_{\mu\nu})$, the gravitational action of the dRGT massive theory of gravity \cite{PhysRevD.82.044020,PhysRevLett.106.231101}, can be re-expressed as \cite{Panpanich:2018}
\begin{equation}
\mathcal{S} = \frac{M_{\mathrm{Pl}}^2}{2}\int\ed x^4\,\sqrt{-g}\Big[
R+m_g^2~\mathcal{U}(\bm{g},\bm{f})
\Big]+\mathcal{S}_\mathrm{m},
    \label{eq:action0}
\end{equation}
in which $\mathcal{S}_\mathrm{m}$ is the matter action, $M_{\mathrm{Pl}}$ represents the reduced Planck mass, $m_g$ is the graviton's mass, and $\mathcal{U}$ is the gravitons' potential, which to avoid the Boulware-Deser ghost, must obey the following self-interaction
\begin{equation}
\mathcal{U} = \mathcal{U}_2+\alpha_3\mathcal{U}_3+\alpha_4\mathcal{U}_4,
    \label{eq:U}
\end{equation}
where
\begin{subequations}
    \begin{align}
        & \mathcal{U}_2=[\mathcal{K}]^2-[\mathcal{K}^2],\\
        & \mathcal{U}_3=[\mathcal{K}]^3-3[\mathcal{K}][\mathcal{K}^2]+2[\mathcal{K}^3],\\
        & \mathcal{U}_4=[\mathcal{K}]^4-6[\mathcal{K}]^2[\mathcal{K}^2]+3[\mathcal{K}^2]^2+8[\mathcal{K}][\mathcal{K}^3]-6[\mathcal{K}^4],
    \end{align}
    \label{eq:U234}
\end{subequations}
in which $[\mathcal{K}]=\mathcal{K}^\mu_\mu$ and $(\mathcal{K}^{i})^\mu_\nu = \mathcal{K}^{\mu}_{\rho_1}\mathcal{K}^{\rho_1}_{\rho_2}\cdots\mathcal{K}^{\rho_i}_{\nu}$, considering the definition 
\begin{equation}
\mathcal{K}^\mu_\nu = \delta^\mu_\nu -\sqrt{g^{\mu\lambda}\partial_\lambda\varphi^{a}\partial_\nu\varphi^b f_{ab}}\,.
    \label{eq:Kmunu}
\end{equation}
In the above expression, it is important to distinguish between the physical metric $\bm{g}$ and the reference metric $\bm{f}$, that both act on the Stückelberg field $\bm{\varphi}$. If the unitary gauge $\bm{\varphi}=\bm{x}\cdot\bm{\delta}$ is taken into account, then one can recast $\sqrt{\bm{g}\cdot\bm{\partial\varphi}\cdot{\bm{\partial\varphi}}\cdot\bm{f}}=\sqrt{\bm{g}\cdot\bm{f}}$ in Eq.~\eqref{eq:Kmunu}. Consequently, the field equations are obtained as
\begin{equation}
G^{\mu}_\nu+m_g^2 X^\mu_\nu = 8\pi G T^{\mu(\mathrm{m})}_\nu,
    \label{eq:fieldEq0}
\end{equation}
in which $\bm{G}$ is the Einstein tensor, $\bm{T}^{(\mathrm{m})}$ is the matter energy-momentum tensor, and \cite{ghosh_class_2016,PhysRevD.85.044024,PhysRevD.87.064001}
\begin{multline}
X^{\mu}_\nu=\mathcal{K}^{\mu}_{\nu}-[\mathcal{K}]\delta^\mu_\nu-\alpha\left[
(\mathcal{K}^2)^\mu_\nu-[\mathcal{K}]\mathcal{K}^\mu_\nu+\frac{1}{2}\delta^\mu_\nu\left([\mathcal{K}]^2-[\mathcal{K}^2]\right)
\right]\\
+3\beta\left[
(\mathcal{K}^3)^\mu_\nu-[\mathcal{K}](\mathcal{K}^2)^\mu_\nu+\frac{1}{2}\mathcal{K}^\mu_\nu\left([\mathcal{K}]^2-[\mathcal{K}^2]\right)
-\frac{1}{6}\delta^\mu_\nu\left([\mathcal{K}]^3-3[\mathcal{K}][\mathcal{K}^2]+2[\mathcal{K}^3]\right)
\right],
    \label{eq:Xmunu}
\end{multline}
is the massive graviton tensor. Now by assuming the ansatz $f_{\mu\nu}=\mathrm{diag}\left(0,0,C^2,C^2\sin^2\theta\right)$ with $C>0$, the terms of order $\mathcal{O}(\mathcal{K}^4)$ are eliminated, and the static spherically symmetric vacuum solution to the dRGT massive gravity will be  characterized by the line element 
\begin{equation}
\ed s^2=-B(r) \ed t^2 + \frac{\ed r^2}{B(r)}+r^2\left(\ed\theta^2+\sin^2\theta\ed\phi^2\right),
    \label{eq:metric0}
\end{equation}
in the usual Schwarzschild coordinates, where the lapse function is given by \cite{Panpanich:2018}
\begin{equation}
    B(r) = 1-\frac{2M}{r}-\frac{1}{3}\Lambda r^2+\gamma r+\zeta,
    \label{eq:lapse}
\end{equation}
in which $\Lambda$ performs the role of the cosmological constant, as in the common Schwarzschild-de Sitter spacetime, and the parameters $\gamma$ and $\zeta$ stem in the massive theory of gravity. The three terms $\{\Lambda,\gamma,\zeta\}$ obey the relationships \cite{Panpanich:2018}
\begin{subequations}
\begin{align}
    & \Lambda = -3m_g^2(1+\alpha+\beta),\label{eq:Lambda}\\
    & \gamma = -m_g^2 C (1+2\alpha+3\beta),\label{eq:gamma}\\
    & \zeta = m_g^2 C^2 (\alpha+3\beta).\label{eq:zeta}
\end{align}
\end{subequations}
The flat space with $\zeta=0$ is obtained in terms of the conditions $\alpha=-3\beta$ and $\beta=1/2+\epsilon$ with $0<\epsilon\ll 1$ \cite{Panpanich:2018}, which is respected in this study to guarantee the positiveness of $\Lambda$ and $\gamma$. 

\begin{figure}[t]
    \centering
    \includegraphics[width=9cm]{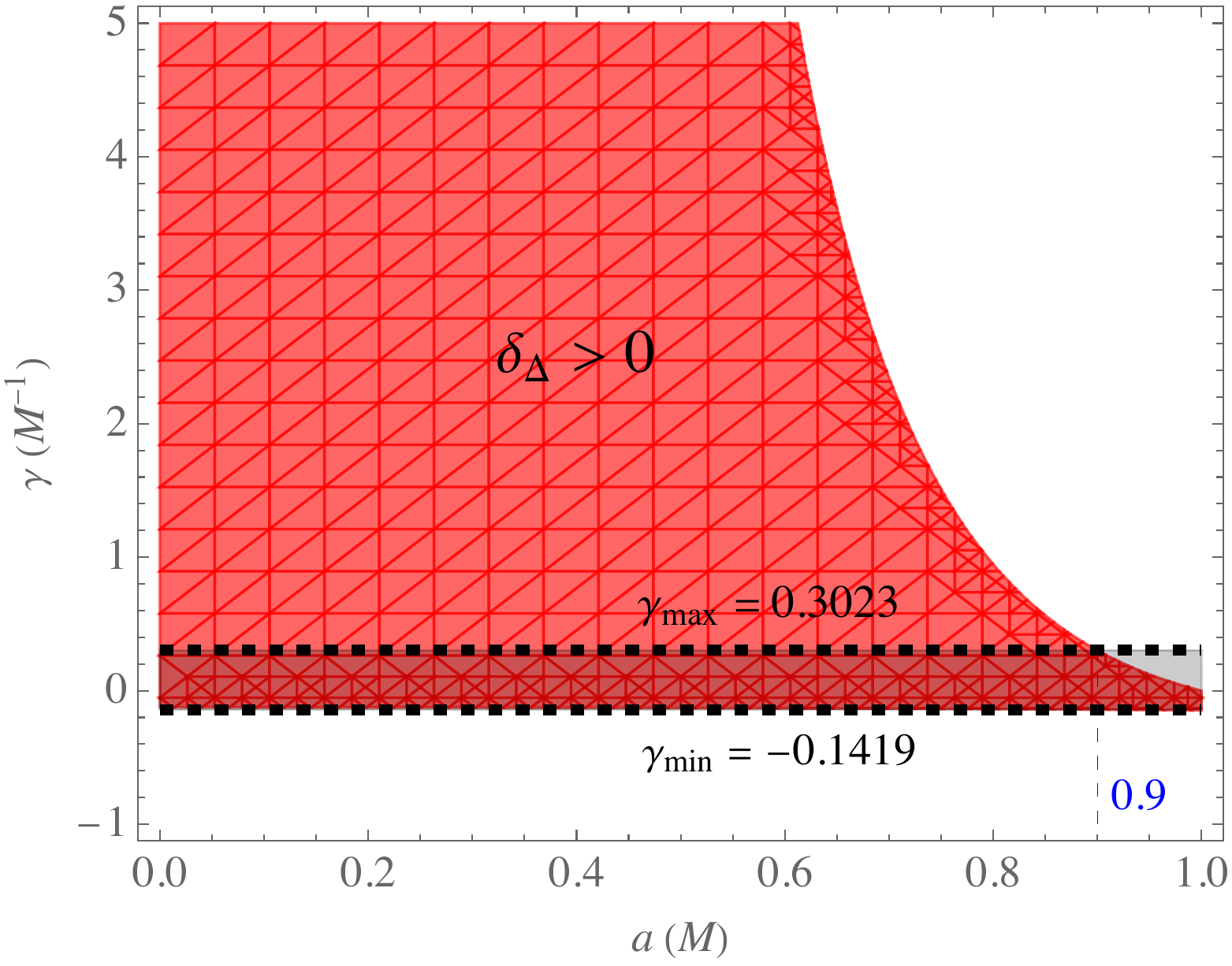}
    \caption{The region plot of $\delta_\Delta>0$ for $\Lambda = 10^{-6}\, M^{-2}$ and $0<a<1$ is shown. According to the diagrams, for each value of $a$, the positivity of the discriminant is guaranteed inside a particular domain $\gamma_{\min}<\gamma<\gamma_{\max}$. In this case, the shaded region represents the domain corresponding to $a=0.9 M$. For all values of $a$, it holds that $\gamma_{\min} < 0$, and the width of the region $\delta_\Delta >0$ significantly increases as the spin parameter decreases. It is worth noting that for the exact values of $\gamma_{\min}$ and $\gamma_{\max}$, $\delta_\Delta=0$, and possessing these values results in extremal black holes.}
    \label{fig:deltaDelta}
\end{figure}

\begin{figure}
    \centering
    \includegraphics[width=7.1 cm]{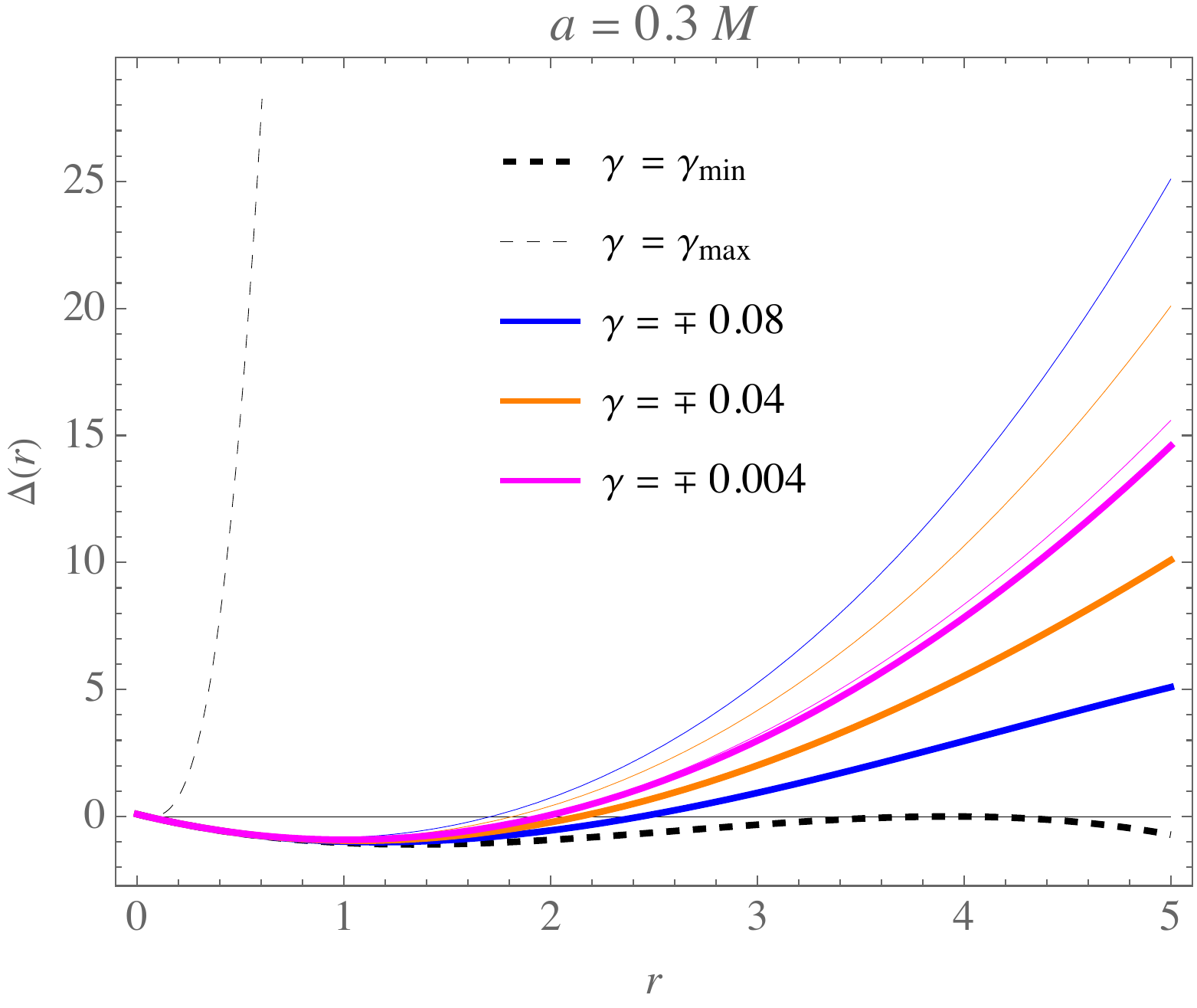}~(a)\qquad
    \includegraphics[width=7.1 cm]{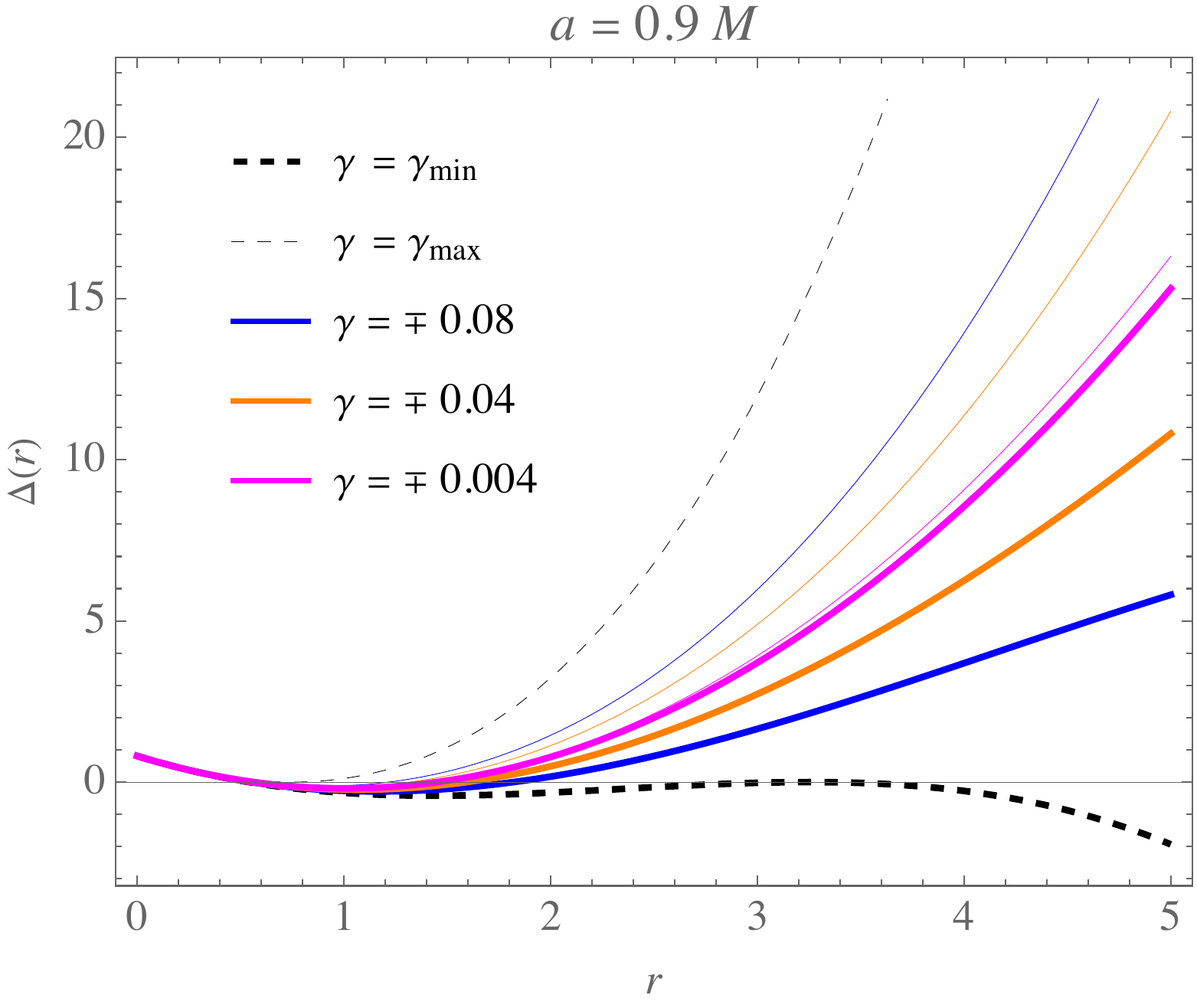}~(b)\qquad
    \caption{The behavior of $\Delta(r)$ for a slow and a fast-rotating black hole, plotted for different values of the $\gamma$-parameter. The thick and thin solid curves represent, respectively, the negative and positive values, and the dashed curves correspond to the extremal cases. }
    \label{fig:Delta}
\end{figure}

\begin{figure}
    \centering
    \includegraphics[width=7.1 cm]{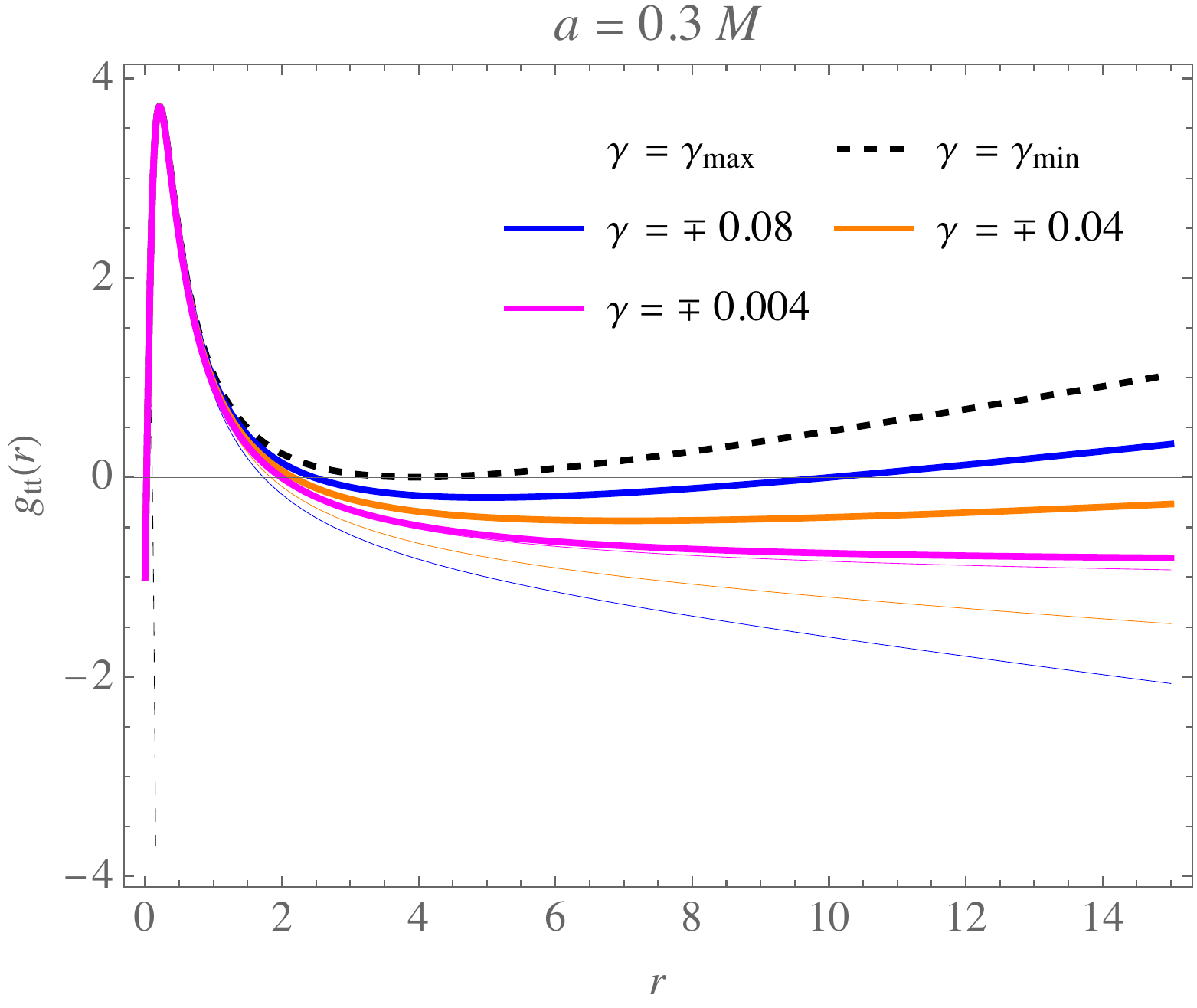}~(a)\qquad
    \includegraphics[width=7.1 cm]{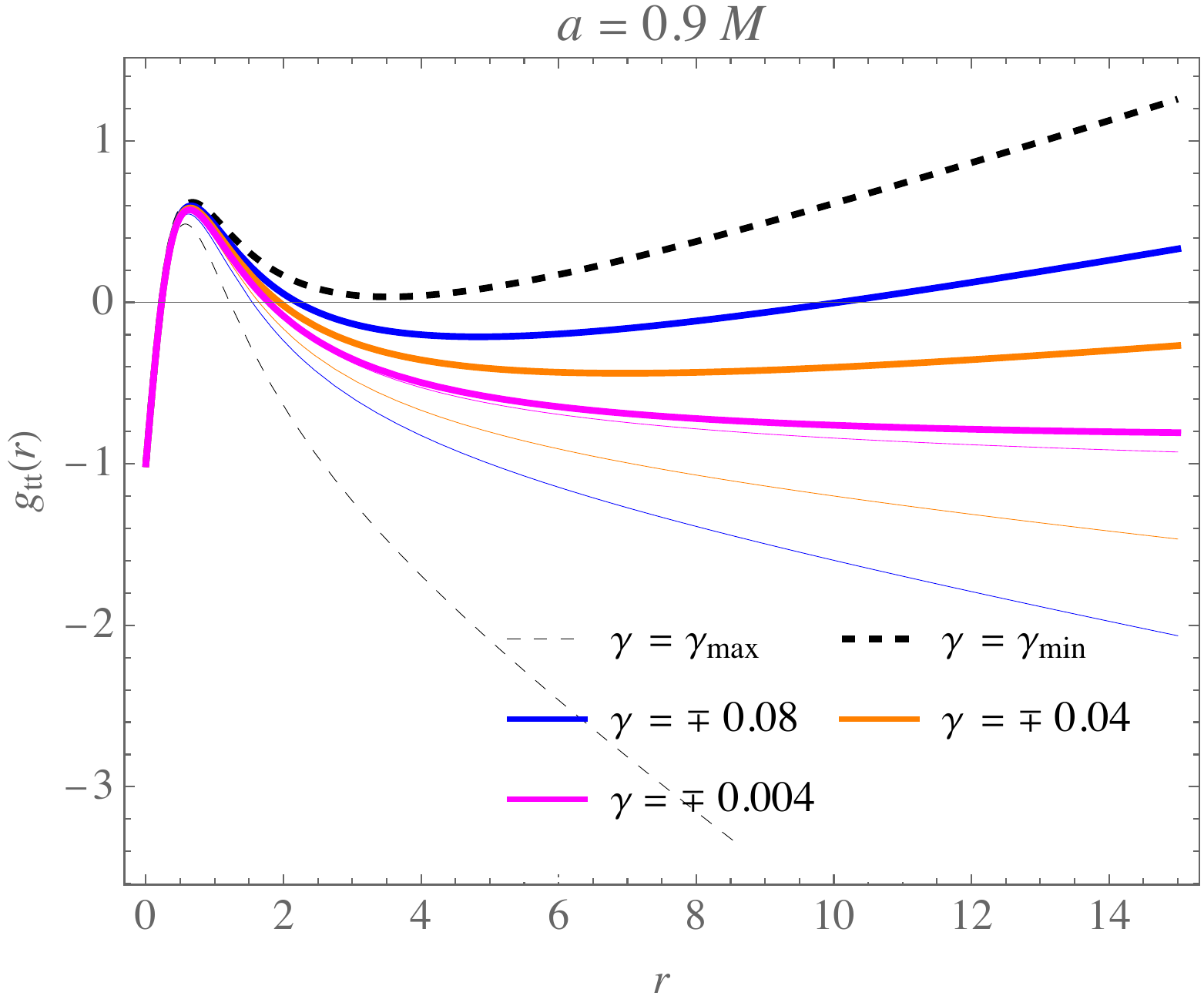}~(b)\qquad
    \caption{The radial profile of $g_{tt}$ plotted for $\theta=\pi/4$ for the two cases of $a=0.3M$ and $a=0.9M$, and for the same values of the  $\gamma$-parameter as in Fig.~\ref{fig:Delta}. The first two roots of $g_{tt} = 0$ are shown for all of the cases, and for the extremal black holes, only two static limits are available. }
    \label{fig:gtt}
\end{figure}

Now, to obtain the rotating counterpart of this black hole spacetime, a modified version of the Newman-Janis algorithm \cite{Newman:1965}, proposed by Azreg-A\"{\i}nou \cite{Azreg:2014} is applied. 
{This method employs a  non-complexification procedure to generate stationary spacetimes from their static counterparts. In order to accomplish this, it is necessary to express the line element \eqref{eq:metric0} using the Eddington-Finkelstein coordinates $(\mathfrak{U},r,\theta,\phi)$. This can be achieved by introducing the transformation $\ed\mathfrak{U}=\ed t-\ed r/B(r)$, which produces 
\begin{equation}
\ed s^2 = -B(r)\ed\mathfrak{U}^2-2\ed\mathfrak{U}\,\ed r+r^2\left(\ed\theta^2+\sin^2\theta\ed\phi^2\right).
    \label{eq:metric0_eddington}
\end{equation}
By introducing the null tetrad set $\mathfrak{Z}_{A}^\mu=\left(l^\mu,n^\mu,m^\mu,\bar{m}^\mu\right)$, in which $\bar{\bm{m}}$ is the complex conjugate of $\bm{m}$, and considering $l^{\mu}=\delta_r^\mu$, $n^\mu=\delta_{\mathfrak{U}}^\mu-\frac{1}{2}B(r)\delta_r^\mu$, and $m^\mu=\big(\delta_\theta^\mu+\mathrm{i}\delta_\phi^\mu/\sin\theta\big)/(\sqrt{2}r)$, one can express the inverse of the metric as $g^{\mu\nu}=-l^\mu n^\nu-l^\nu n^\mu+m^\mu\bar{m}^\nu+m^\nu\bar{m}^\mu$. Note that, the tetrad vector fields obey the conditions $\bm{l}\cdot\bm{l}=\bm{n}\cdot\bm{n}=\bm{m}\cdot\bm{m}=\bm{l}\cdot\bm{m}=\bm{n}\cdot\bm{m}=0$, and $\bm{l}\cdot\bm{n}=-\bm{m}\cdot\bar{\bm{m}}=-1$. Now to incorporate the spin parameter $a$ of the stationary counterpart, we apply the transformation $\delta_\theta^\mu\rightarrow\delta_r^\mu+\mathrm{i}a\sin\theta(\delta_\mathfrak{U}^\mu-\delta_r^\mu)$, while all the others remain unchanged \cite{Azreg:2014}. This procedure leads to the transformations $B(r)\rightarrow \mathfrak{B}(r,a,\theta)$ and $r^2\rightarrow \mathfrak{H}(r,a,\theta)$. Consequently, the null tetrad assumes the form
\begin{subequations}
    \begin{align}
        & l'^\mu=\delta_r^\mu,\\
        & n'^\mu=\delta_\mathfrak{U}^\mu-\frac{\mathfrak{B}(r,a,\theta)}{2}\delta_r^\mu,\\
        & m'^\mu=\frac{1}{\sqrt{2\mathfrak{H}(r,a,\theta)}}\left[
        \mathrm{i}a\sin\theta\left(\delta_{\mathfrak{U}}^\mu-\delta_r^\mu\right)+\delta_\theta^\mu+\frac{\mathrm{i}}{\sin\theta}\delta_\phi^\mu
        \right],
    \end{align}
   \label{eq:tetradprime}
\end{subequations}
and the new inverse metric becomes $g^{\mu\nu}=-l'^\mu n'^\nu-l'^\nu n'^\mu+m'^\mu\bar{m}'^\nu+m'^\nu\bar{m}'^\mu$ in the Eddington-Finkelstein coordinates. To transition to the desired Boyer-Lindquist coordinates, we adopt the transformations $\ed\mathfrak{U}=\ed t'+\omega(r)\ed r$ and $\ed\phi=\ed\phi'+\chi(r)\ed r$, and we choose \cite{Azreg:2014}
\begin{subequations}
    \begin{align}
       & \omega(r)=-\frac{r^2+a^2}{r^2B(r)+a^2},\\
       & \chi(r) = -\frac{a}{r^2B(r)+a^2}.
    \end{align}
    \label{eq:omega,chi}
\end{subequations}
The remaining steps involve setting $\mathfrak{B}(r,a,\theta)=\mathfrak{H}(r,a,\theta)^{-2}[r^2B(r)+a^2\cos^2\theta]$ in order to eliminate the cross term $\ed t\ed r$ in the line element. Additionally, we select $\mathfrak{H}(r,a,\theta)=r^2+a^2\cos^2\theta$ to satisfy $G_{r\theta}=0$. Finally,
}
this algorithm generates the stationary spacetime of the rotating dRGT (termed as RdRGT) black hole in the form
\begin{multline}\label{eq:metric_rotating}
    \ed s^2 = -\frac{\Delta-a^2\sin^2\theta}{\Sigma} \ed t^2 + \frac{\Sigma}{\Delta} \ed r^2 - 2 a \sin^2\theta\left(1-\frac{\Delta-a^2\sin^2\theta}{\Sigma}\right)\ed t\ed \phi + \Sigma \ed\theta^2\\
    + \sin^2\theta\left[\Sigma+a^2\sin^2\theta\left(2-\frac{\Delta-a^2\sin^2\theta}{\Sigma}\right)\right]\ed\phi^2,
\end{multline}
in which 
the spin parameter is given by $a = J/M$, with $J$ being the black hole's angular momentum. We have also defined
\begin{subequations}
\begin{align}
    & \Delta(r) = a^2 + r^2 B(r) = \gamma r^3+
    r^2+a^2 - 2Mr - \frac{\Lambda}{3}r^4,\label{eq:Delta}\\
    & \Sigma(r,\theta) = r^2+a^2 \cos^2\theta.
\end{align}
\end{subequations}
The stationary black holes are surrounded by several hyper-surfaces that configure their exterior spacetime. First of all, the horizons of the RdRGT black hole are located by the roots of $\Delta(r)=0$, which results in the quartic equation
\begin{equation}
\Lambda r^4-3r^2\left(1+\gamma r\right)+6M r-3a^2 = 0,
    \label{eq:Delta=0}
\end{equation}
admitting the discriminant
\begin{multline}
\delta_{\Delta} = \frac{1}{27} \left(-256 a^6 \Lambda ^3-729 a^4 \gamma ^4-1296 a^4 \gamma ^2 \Lambda -384 a^4 \Lambda ^2+1152 a^4 \gamma  \Lambda ^2 M-108 a^2 \gamma ^2\right.\\
\left.-144 a^2 \Lambda +216 a^2 \gamma ^2 \Lambda  M^2+1728 a^2 \Lambda ^2 M^2-972 a^2 \gamma ^3 M-1440 a^2 \gamma  \Lambda  M-1296 \Lambda ^2 M^4\right.\\
\left.+864 \gamma ^3 M^3+1296 \gamma  \Lambda  M^3+108 \gamma ^2 M^2+144 \Lambda  M^2\right).
    \label{eq:DisDelta}
\end{multline}
Since $\Lambda\ll1$, the above discriminant can be positive within a certain range for the values of $\gamma$, and hence, the above quartic can possess four real roots, one negative and three positives. In Fig. \ref{fig:deltaDelta}, the region inside which, $\delta_\Delta > 0$, has been presented when the black hole's spin parameter varies from zero to one. Accordingly, for $\gamma_{\min}<\gamma<\gamma_{\max}$, the positivity of the discriminant is guaranteed for each value of $a$, and the cases of $\gamma = \gamma_{\min}$ and $\gamma = \gamma_{\max}$ correspond to extremal black holes. 
The figure shows that the width of this domain decreases with the increase of the spin parameter. Therefore, the condition $\delta_\Delta=0$ is satisfied for both fast and slow-rotating black holes if we choose $|\gamma|\ll1$. Note that, the extremal limits $\gamma_{\substack{{\min}\\{\max}}}$ correspond to a vanishing discriminant, for which the equation $\Delta(r)=0$ has one negative and three positive roots, which two of them are degenerate (coinciding horizons). On the other hand, the equation $\delta_\Delta=0$ has one positive, one negative, and two complex conjugate roots, and hence, one can recast the discriminant in Eq. \eqref{eq:DisDelta} as $\delta_\Delta = 27 a^4 (\gamma_{\max}-\gamma)(\gamma-\gamma_{\min})(\gamma^2-|\gamma_1|^2)$, in which $\gamma_1\in\Bbb{C}$. This quartic equation can be solved explicitly for $\gamma_{\substack{{\min}\\{\max}}}$, but we find it unnecessary to present their analytical expressions. On the other hand, the suppressed quartic \eqref{eq:Delta=0} has the four analytical roots {$r_1, r_2, r_3$, and $r_4$, as they have been presented in appendix \ref{app:A}. It can be verified that for the aforementioned domain of $\gamma$, the solutions \eqref{eq:r1}--\eqref{eq:r4} admit $r_4<0$ and $r_1>r_2>r_3>0$.} This way, the black hole will possess three horizons, which from smaller to larger are namely, the Cauchy horizon $r_-=r_3$, the event horizon $r_+=r_2$, and the cosmological horizon $r_{++}=r_1$. Once $\gamma = \gamma_{\max}$, the extremal black hole is obtained as the Cauchy and event horizons coincide, whereas for the case of $\gamma = \gamma_{\min}$, extremality corresponds to the unification of the event and cosmological horizons. In Fig. \ref{fig:Delta}, the behavior of $\Delta(r)$ has been plotted for different values of the $\gamma$-parameter, where the solutions to $\Delta=0$ and the extremal limit have been indicated.
The other hypersurfaces that characterize the stationary spacetimes are those that correspond to the static limits, and together, they form the black holes' ergoregions. These regions, inside which no static observer can exist, are identified by the equation $g_{tt} = 0$, in accordance with the line element \eqref{eq:metric_rotating}. For the RdRGT black hole this results in another quartic, whose solutions are exactly the same positive solutions $r_{\mathrm{st}_{++}} = r_1$, $r_{\mathrm{st}_+} = r_2$ and $r_{\mathrm{st}_-} = r_3$, as those given in Eqs. \eqref{eq:r1}--\eqref{eq:r4} with the same form for the included parameters, considering the only replacement $a^2\rightarrow a^2\cos^2\theta$ in Eq.~\eqref{eq:mathcalC}. Hence, it is straightforward to see that for $\theta = 0$, the static limits and the black hole horizons coincide. Also as expected, these radii satisfy the conditions
$0<r_{\mathrm{st}_-}<r_-$ and $r_+<r_{\mathrm{st}_+}<r_{\mathrm{st}_{++}}<r_{++}$, and comprise the three boundaries of the interior and exterior ergoregions. In Fig.~\ref{fig:gtt}, the radial profile of the function $g_{tt}$ has been plotted for the same parameters as in Fig.~\ref{fig:Delta}.
{To focus on the motion of particles in the exterior geometry of the RdRGT black hole, we limit our analysis to the region defined by $r_+<r<r_{++}$. Within this domain, we utilize Carter's geodesic equations of motion to study the behavior of particles in orbits with constant radii.}

\section{Spherical particle orbits}\label{sec:geodesic}

Black holes with non-zero spin parameter can form several spherical particle trajectories around them, each of which, stays on a sphere with $r=\mathrm{const.}$, when the $\theta$-coordinate oscillates between two turning points (for the case of Kerr black holes, see for example Refs. \cite{Bardeen:1972a,Bardeen:1973b,Chandrasekhar:1998}). These spherical orbits exist in a certain interval and only a subclass of them, such as the stable circular orbits that are confined by the ISCO and OSCO are on the equatorial plane, while all the other orbits are non-planar. In this section, we employ the standard geodesic equations for massive particles to investigate the planar and non-planar $r$-constant particle orbits around the RdRGT black hole.

Applying the Carter's method of separation of the Hamilton-Jacobi equation \cite{Carter:1968}, we can determine the time-like geodesic equations in the exterior geometry of the RdRGT black hole, which are given as
\begin{eqnarray}
M\frac{\ed r}{\ed\lambda} &=&\pm\sqrt{\mathcal{R}(r)},\label{eq:drdtau}\\
M\frac{\ed \theta}{\ed\lambda} &=&\pm\sqrt{\Theta(\theta)},\label{eq:dthetadtau}\\
M\frac{\ed \phi}{\ed\lambda} &=&\frac{a}{\Delta}\left[E\left(r^2+a^2\right)-aL\right]-\left(aE-\frac{L}{\sin^2\theta}\right),\label{eq:dphidtau}\\
M\frac{\ed t}{\ed\lambda} &=& \frac{r^2+a^2}{\Delta}\left[E\left(r^2+a^2\right)-aL\right]-a\left(aE\sin^2\theta-L\right),\label{eq:dtdtau}
\end{eqnarray}
where we have used the dimensionless \textit{Mino time}, $\lambda$, defined as $\Sigma\ed\lambda=M\ed\tau$ \cite{Mino:2003}, in which $\tau$ is the geodesics' affine parameter. In Eqs.\eqref{eq:drdtau}--\eqref{eq:dtdtau}, $E$ and $L$ are the constants of motion related to the Killing symmetries of the spacetime, which are termed, respectively, as the energy and angular momentum of the particles. Furthermore, 
\begin{subequations}
\begin{align}
&     \mathcal{R}(r)= \Big[E\left(r^2+a^2\right)-aL\Big]^2-\Delta\Big[\epsilon^2 r^2+\left(L-aE\right)^2+\Q\Big],\label{eq:R}\\
& \Theta(\theta) = \Q-\cos^2\theta\left[a^2\left(\epsilon^2-E^2\right)+\frac{L^2}{\sin^2\theta}\right],\label{eq:V}
\end{align}
\end{subequations}
defining the Carter's constant $\Q$, as the third constant of motion. Since we are interested in the motion of massive particles, we set $\epsilon=1$. Note that, from Eqs.~\eqref{eq:drdtau} and \eqref{eq:dthetadtau}, one can infer that the particles' motion is performed under the mutual conditions $\mathcal{R}(r)\geq 0$ and $\Theta(\theta)\geq 0$. Also for the sake of convenience, the positive segments of these equations are adopted. 
It is important to accentuate the role of $\Q$ in the classification of the orbits; orbits on the equatorial plane (planar orbits that correspond to $\theta=\pi/2$) are obtained for $\Q=0$. 
These orbits can be regarded as the boundaries of the spherical orbits in their general case, for which, $\Q\geq0$. In fact, circular orbits satisfy the conditions $\mathcal{R}(r)=0=\mathcal{R}'(r)$, which provides the critical quantities
\begin{multline}
L_c(r) = \frac{1}{3 a \Big[6 M+r \left(4 \Lambda  r^2-9 \gamma  r-6\right)\Big]}\Bigg\{3 E \bigg[a^2 \bigg(6 M+r^2 (4 \Lambda  r-9 \gamma )\bigg)-6 M r^2+r^4 (2 \Lambda  r-3 \gamma )\bigg]\\
- \bigg[\Lambda  r^4-3 \bigg(a^2+r \left(r-2 M+\gamma  r^2\right)\bigg)\bigg]\sqrt{6r  \bigg[6 M+r \left(4 \Lambda  r^2-9 \gamma  r-6\right)+6 E^2 r\bigg]}\,
\Bigg\},
\label{eq:L_c}
\end{multline}
\begin{multline}
\Q_c(r) = \frac{r^2}{3 a^3 \Big[6 M+r \left(4 \Lambda  r^2-9 \gamma  r-6\right)\Big]^2}\Bigg\{a r \Bigg[27 E^2 r \bigg[20 M^2-4 M r (3 \gamma  r+4)+r^2 \Big(\gamma  r (5 \gamma  r+8)+4\Big)\bigg]\\
+6 \Lambda ^2 r^6 \Big[18 M-r (11 \gamma  r+10)+4 E^2 r\Big]+36 \Lambda  r^3 \bigg[E^2 r\Big(2 M-r (3 \gamma  r+2)\Big)+\Big(2 M-r (\gamma  r+1)\Big)\\
\times\Big(6 M-r (5 \gamma  r+4)\Big)\bigg]
+54 \left(r-2 M+\gamma  r^2\right)^2 \Big(2 M-r (3 \gamma  r+2)\Big)+8 \Lambda ^3 r^9\Bigg]\\
+3 a^3 \Big(6 M+r^2 (3 \gamma -2 \Lambda  r)\Big) \Big(6 M+r \left(4 \Lambda  r^2-9 \gamma  r-6\right)+12 E^2 r\Big)+2 \sqrt{6} E \Big(6 M+r^2 (3 \gamma -2 \Lambda  r)\Big)\\
\times\bigg[3 a^3+a r \Big(r \left(3-\Lambda  r^2+3 \gamma  r\right)-6 M\Big)\bigg]\sqrt{r  \Big(6 M+r \left(4 \Lambda  r^2-9 \gamma  r-6\right)+6 E^2 r\Big)}
\Bigg\}.
\label{eq:Q_c}
\end{multline}
Furthermore, the effective inclination angle can be defined by means of the relation \cite{Tavlayan:2021}
\begin{equation}
\cos i = \frac{L}{\sqrt{L^2+\Q}}.
    \label{eq:cosi}
\end{equation}
Using this relation, {and by eliminating $L$ between the equations $\mathcal{R}(r)=0$ and $\mathcal{R}'(r)=0$, we get to the single equation}
\begin{equation}
p_{14}(x)=\sum_{j=0}^{14}m_j x^j=0,
    \label{eq:p14}
\end{equation}
as the general characteristic equation for the radii of spherical particle orbits in the RdRGT spacetime, where
\begin{subequations}
\begin{align}
 & m_0 = 36 k^2 \nu ^2 u^4,\\
 & m_1 = 72 \left(k-1\right) k \nu ^2 u^4 ,\\
 & m_2 = 36 \nu  u^3 \left(-4 k^2 \nu +k^2 \nu  u+10 k \nu -2 k \nu  u-2 k+\nu  u\right),\\
 & m_3 = 72 \nu  u^2 \left(2 k^2 \nu  u-2 k^2 u-8 k \nu +k \nu  u+3 k u+4 k-3 \nu  u-u\right),\\
 & m_4 = 36 u^2 \left(-5 h k \nu ^2 u-12 k^2 \nu ^2+10 k^2 \nu +4 k^2 \nu ^2 u+12 k \nu ^2-28 k \nu -8 k \nu ^2 u+9 \nu ^2+14 \nu +4 \nu ^2 u+1\right),\\
 & m_5 = -36 u \left(5 h k \nu ^2 u^2-16 h k \nu ^2 u+8 h k \nu  u-5 h \nu ^2 u^2+12 k^2 \nu  u-2 k l \nu ^2 u^2-16 k \nu -12 k \nu ^2 u-24 k \nu  u+24 \nu\right.\nonumber\\
 &\left.\qquad+12 \nu ^2 u+12 \nu  u+8\right),\\
 & m_6 = 12 \left(45 h k \nu  u^2-45 h \nu ^2 u^2-39 h \nu  u^2+12 k^2 \nu ^2 u^2+6 k^2 \nu  u^2+36 k^2 \nu  u+6 k l \nu ^2 u^3-20 k l \nu ^2 u^2+10 k l \nu  u^2\right.\nonumber\\
&\left.\qquad-24 k \nu ^2 u^2-12 k \nu  u^2-138 k \nu  u-30 k u-6 l \nu ^2 u^3+12 \nu ^2 u^2+6 \nu  u^2+120 \nu  u+24 u+48\right),\\
& m_7 = -12 \left(30 h k \nu ^2 u^2+9 h k \nu  u^2+48 h k \nu  u-30 h \nu ^2 u^2-9 h \nu  u^2-132 h \nu  u-24 h u+36 k^2 \nu  u+12 k^2 u\right.\nonumber\\&
\left.\qquad+18 k l \nu  u^2-102 k \nu  u-18 k u-72 k-18 l \nu ^2 u^2-16 l \nu  u^2+66 \nu  u+6 u+96\right),\\
& m_8 = 3 \left(75 h^2 \nu ^2 u^2+372 h k \nu  u+72 h k u-540 h \nu  u-36 h u-384 h+48 k^2 \nu  u+108 k^2+48 k l \nu ^2 u^2+16 k l \nu  u^2\right.\nonumber\\
&\left.\qquad+80 k l \nu  u-96 k \nu  u-384 k-48 l \nu ^2 u^2-16 l \nu  u^2-216 l \nu  u-40 l u+48 \nu  u+288\right),\\
& m_9 = -12 \left(60 h^2 \nu  u+33 h k \nu  u+72 h k+15 h l \nu ^2 u^2-33 h \nu  u-132 h+18 k^2+38 k l \nu  u+8 k l u-42 k-56 l \nu  u\right.\nonumber\\
&\left.\qquad-4 l u-40 l+24\right),\\
& m_{10} = 6 \left(45 h^2 \nu  u+96 h^2+102 h k+98 h l \nu  u-120 h+6 k^2+28 k l \nu  u+60 k l-12 k+6 l^2 \nu ^2 u^2-28 l \nu  u-112 l\right.\nonumber\\
&\left.\qquad+6\right),\\
& m_{11} = -12 \left(36 h^2+9 h k+19 h l \nu  u+40 h l-9 h+22 k l+10 l^2 \nu  u-26 l\right),\\
& m_{12} = 81 h^2+372 h l+48 k l+48 l^2 \nu  u+100 l^2-48 l,\\
& m_{13} = -8 l (9 h+10 l),\\
& m_{14} = 16 l^2,
\end{align}
\label{eq:mj}
\end{subequations}
in which we have used the dimensionless quantities 
\begin{subequations}
\begin{align}
    & x = \frac{r}{M},\\
    & u = \frac{a^2}{M^2},\\
    & k = E^2,\\
    & l = \frac{1}{3}\Lambda M^2,\\
    & h = \gamma M,
\end{align}
\label{eq:x,u,l,h}
\end{subequations}
and the definition $\nu=\sin^2 i$, for which, the Carter's constant takes the form \cite{Tavlayan:2021}
\begin{equation}
\Q = \frac{\nu L^2}{1-\nu}.
    \label{eq:Q_new}
\end{equation}
In general, the polynomial equation \eqref{eq:p14} has fourteen solutions of the form $x(u,\nu,k,h,l)$, among which, those real roots that reside in the domain $x_+<x<x_{++}$ are of our interest. On the other hand, and in accordance with the Abel-Ruffini theorem, this equation cannot be solved analytically in terms of finite radicals. We, therefore, consider some specific cases in what follows in the rest of this section. Note that, for the case of Kerr black hole with $h=l=0$, the above equation reduces to a polynomial equation of order ten, whose analytical solutions to the radii of spherical time-like orbits has been investigated in Ref.~\cite{Tavlayan:2021}. 

One can also make a primary classification of orbits, based on the value of the $k$-parameter. In this regard, for $0\leq k<1$, the test particles travel on \textit{bound orbits}, and accordingly, the case of $k=1$ corresponds to \textit{marginally bound orbits}. On the other hand, $k>1$ provides \textit{unbound orbits}. For unbound spherical orbits the particles travel on unstable trajectories that are deflected from the black hole upon small radial perturbations. This is while for the bound orbits, the particle trajectories can still be stable or unstable. In the latter case, the mentioned perturbations will generate eccentric orbits that oscillate between two finite radial distances \cite{teo_spherical_2021}.

To proceed with our discussion, we restrict ourselves to the case of planar 
orbits at the vicinity of the black hole's event horizon.

\subsection{Radii of planar orbits}\label{eq:xplanar}

\begin{figure}[t]
    \centering
    \includegraphics[width=8cm]{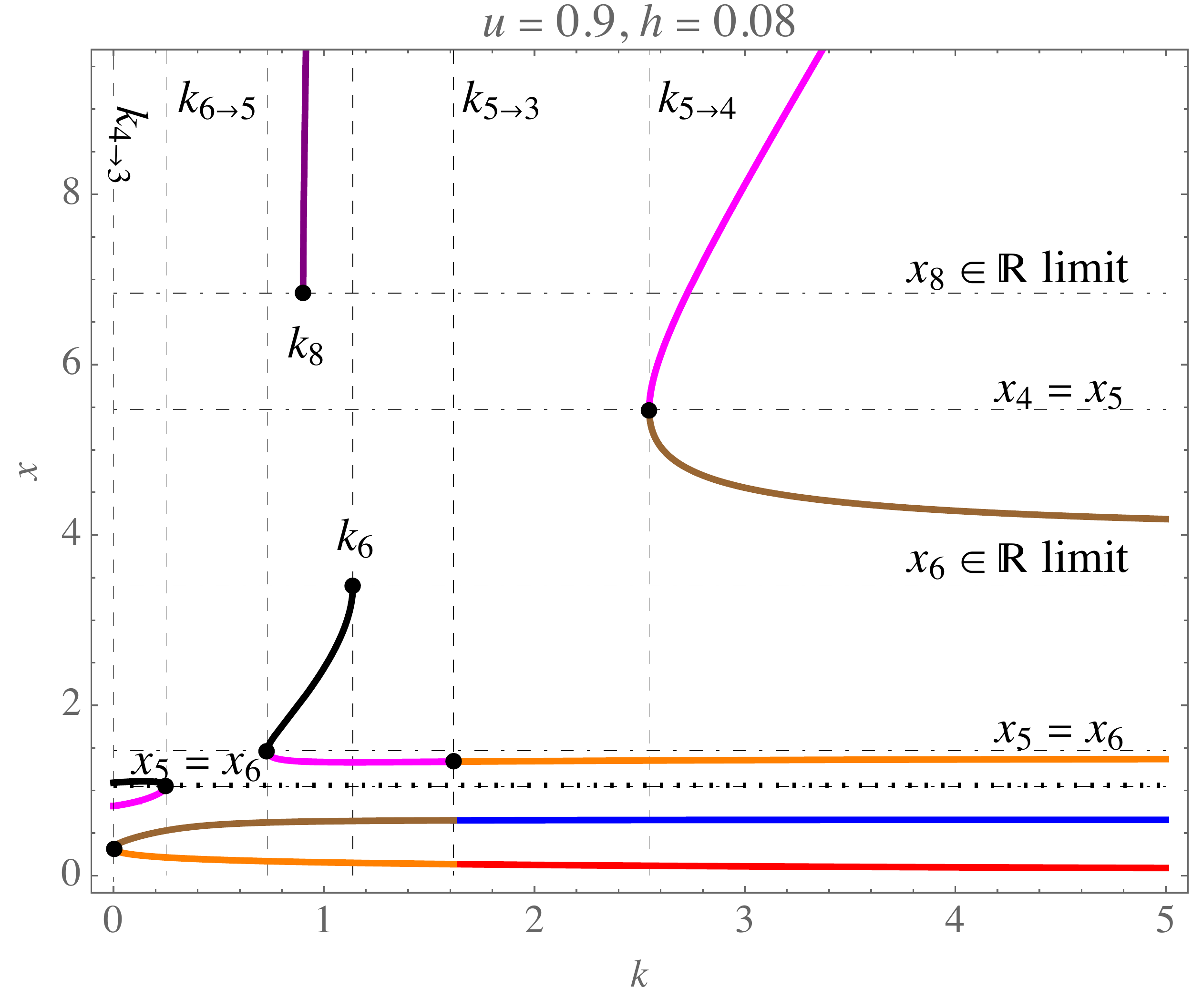}
    \caption{The $x-k$ diagram for the real parts of the solutions $x_i$, plotted for $h=0.08$, $l=10^{-6}$, and $u=0.9$.}
    \label{fig:x-k0.9}
\end{figure}

Planar orbits correspond to trajectories that occur in the equatorial plane, for which $\Q = 0$ (i.e., $i = 0$, $\pi$, or $\nu = 0$). These orbits are of great importance in black hole astrophysics since they can also characterize the orbit of particles within the accretion disk which are confined within the boundaries composed by the ISCO and OSCO. The above condition reduces the order of the polynomial equation \eqref{eq:p14} down to ten. A further reduction relies on the smallness of $h$ and $l$, and their negligible impacts at the vicinity of the black hole event horizon. In this sense, we can take into account only up to the first order of these parameters for our purpose. Hence, the characteristic polynomial equation reduces to the nonic equation.
\begin{equation}
p_9(x)=x^9+\sum_{j=0}^{8}\bar{m}_j x^j=0,
    \label{eq:p9}
\end{equation}
in which
\begin{subequations}
\begin{align}
    & \bar{m}_0 = -\frac{u^2}{2 h l},\\
    & \bar{m}_1 = \frac{4 u }{h l},\\
    & \bar{m}_2 = \frac{ (5 k-4) u-8}{ h l},\\
    & \bar{m}_3 = \frac{16- u \left(4 h-2 k^2+3 k-1\right)-12 k}{ h l},\\
    & \bar{m}_4 = \frac{3 \left(32 h-9 k^2+32 k-24\right)+u (10 l-18 h k+9 h)}{6 h l},\\
    & \bar{m}_5 = \frac{ \left(36 h k-66 h+9 k^2-21 k-20 l+12\right)+2 (2 k-1) l u}{3 h l},\\
    & \bar{m}_6 = \frac{-51 h k+60 h-3 k^2-30 k l+6 k+56 l-3}{6 h l},\\
    & \bar{m}_7 = \frac{9 h k+40 h l-9 h+22 k l-26 l}{6 h l}
    ,\\
    & \bar{m}_8 = \frac{-31 h-4 k+4}{6 h}.
\end{align}
\label{eq:barmj}
\end{subequations}
To study the roots of this equation, $x(u,k,h,l)$, let us first fix $u=u_0$, $h=h_0$, and $l=l_0$. This way, the solutions will be of the form $x_i(k)$ $(i=\overline{1,9})$, for which, we have presented some examples in the following.
\begin{figure}[t]
    \centering
    \includegraphics[width=8cm]{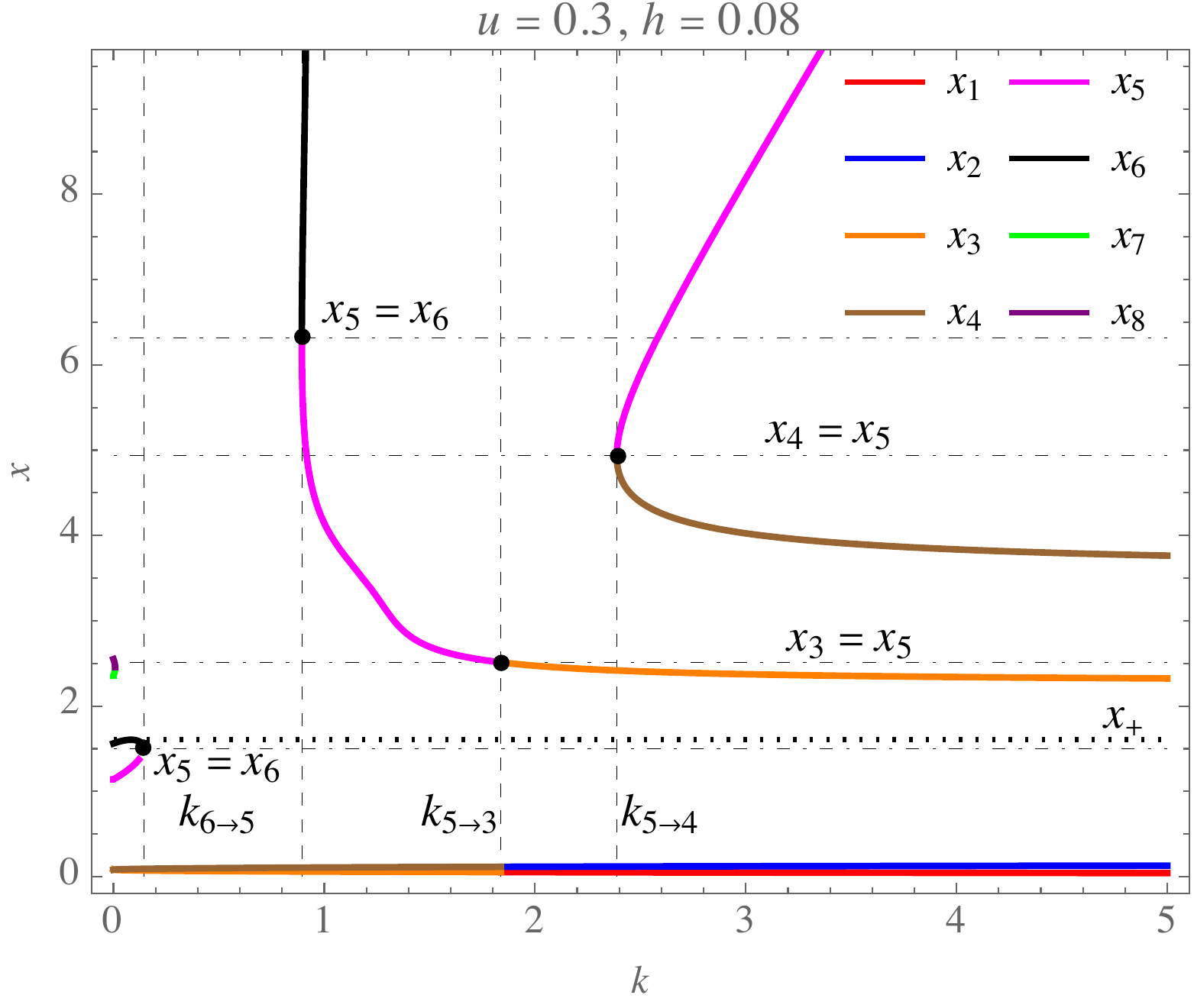}~(a)\qquad
    \includegraphics[width=8cm]{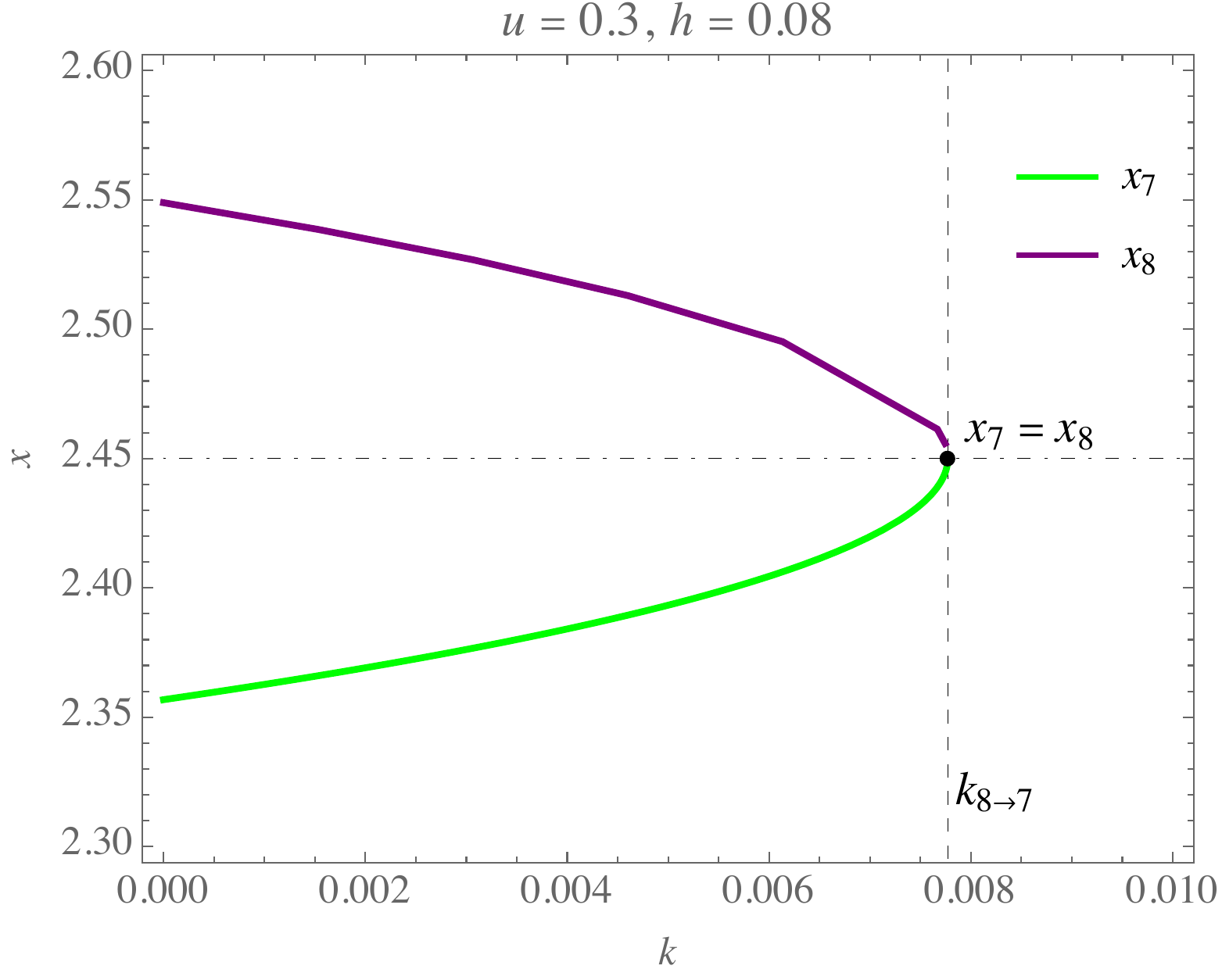}~(b)
    \caption{The $x-k$ diagrams for the real parts of the solutions $x_i$  of Eq. \eqref{eq:p9}, plotted for $h=0.08$, $l=10^{-6}$, and $u=0.3$. In panel (a), the whole range of $k$ has been shown, whereas the range at which $x_7$ and $x_8$ are real, has been magnified in panel (b). The color-coding of the solutions $x_i$ that is used here will be also applied in all of the forthcoming diagrams within the paper.}
    \label{fig:x-k0.3}
\end{figure}
In Fig.~\ref{fig:x-k0.3}, the $k$-profiles of the solutions within the domains where $x_i\in\mathbb{R}$, have been plotted for a slowly rotating black hole with $u_0=0.3$, $h_0=0.08$ and $l_0=10^{-6}$, for which $x_+=1.607$. In these diagrams, by $k_{i\rightarrow j}$ we notate the values of $k$, at which, the solutions $x_i$ and $x_j$ $(i\leq j)$ coincide. As it is seen in the left panel of the figure, $x_3,x_4<x_+$ for $0<k<k_{5\rightarrow3}$. After this point, they are of real values, only for $x_3,x_4>x_+$. Both of these radii connect to $x_5$, respectively, after $k_{5\rightarrow3}=1.838$ and $k_{5\rightarrow4}=2.388$, and at $x=2.512$ and $x=4.934$. Furthermore, $x_1$ and $x_2$ are real-valued only inside the event horizon. $x_5$ has two significant branches in the domains $k_{6\rightarrow5}=0.894<k<k_{5\rightarrow3}$ and $k>k_{5\rightarrow4}$, that respect the condition $x_5>x_+$. At its upper limit,  $x_5$ connects $x_6$ at $x=6.318$. Note that, these radii correspond to unbound orbits. In fact, bound orbits can happen inside the event horizon for the case of $x_5,x_6<x_+$, within the domain $0<k<k_{6\rightarrow5}=0.144$. These two branches coincide at $x=1.502$.  The other possibility corresponds to the solutions $x_7$ and $x_8$, which have been shown, separately, in the right panel of Fig.~\ref{fig:x-k0.3}. These radii are located outside the event horizon and are real-valued in the domain $0<k<k_{8\rightarrow7}=0.008$, for which, they coincide at $x=2.45$. Since this latter case corresponds to bound orbits outside the event horizon, it is worth discussing the stability of the orbits in these radii, which will be referred to further in this subsection. But before that, let us consider the $k$-profile of the solutions for a fast-rotating black hole with $u=0.9$ and the same values for the other parameters, for which, $x_+=1.603$. In Fig.~\ref{fig:x-k0.9}, the regions where $x_i\in\Bbb{R}$ have been shown for this particular case. 
The same as before, the real parts of $x_8$ and $x_9$ coincide. According to the figure, $x_3,x_4<x_+$ for $k_{4\rightarrow3}=0.0003<k<k_{5\rightarrow3}=1.615$. Beyond this domain, $x_3$ and $x_4$ switch, respectively, to $x_1$ and $x_2$ inside the event horizon, and outside it, they have real values after $k_{5\rightarrow3}$ and $k_{5\rightarrow4}=2.546$. The solution $x_5$ has three branches which are ramified as follows. For $k_{4\rightarrow3}<k<k_{6\rightarrow5}=0.250$, it is $x_5<x_+$, whereas for $k_{6\rightarrow5}=0.731<k<k_{5\rightarrow4}$, it holds $x_5>x_+$. A particular domain for this case corresponds to $0.731<k<k_8=0.901$, within which, bound orbits can form outside the event horizon. In this domain, the lower limit of $x_8\in\Bbb{R}$ occurs at $x=6.840$. Beyond this point, $x_5>x_+$ is real-valued only in the region $k>k_{5\rightarrow4}$. The solution $x_6$ is divided into two branches outside the event horizon, which correspond to the domains $k_{4\rightarrow3}<k<k_{6\rightarrow5}=0.250$ and $k_{6\rightarrow5}=0.731<k<k_6=1.137$, within which, the upper limit of $x_6\in\Bbb{R}$ occurs at $x=3.40$.  Again, bound orbits can form inside the domain $0.731<k<k_8$. Note that, the real part of the solution $x_7$ coincides with that of $x_6$. 

To proceed further, it is also important to distinguish between the directions of planar orbits outside the event horizon, within the energy domains discussed above. For this reason, we construct the set of solutions $x_i(u)\equiv x(u,k_0,h_0,l_0)$, so that the $u$-profiles of the roots of the nonic equation can be checked. Considering the previous values for $h_0$ and $l_0$, the critical value of the spin parameter is $u_\mathrm{ext}=0.923$, for which, $x_+=x_-=x_{\mathrm{ext}}=0.902$, giving the extremal black hole horizon where the event and Cauchy horizons coincide. 
\begin{figure}[t]
    \centering
    \includegraphics[width=8cm]{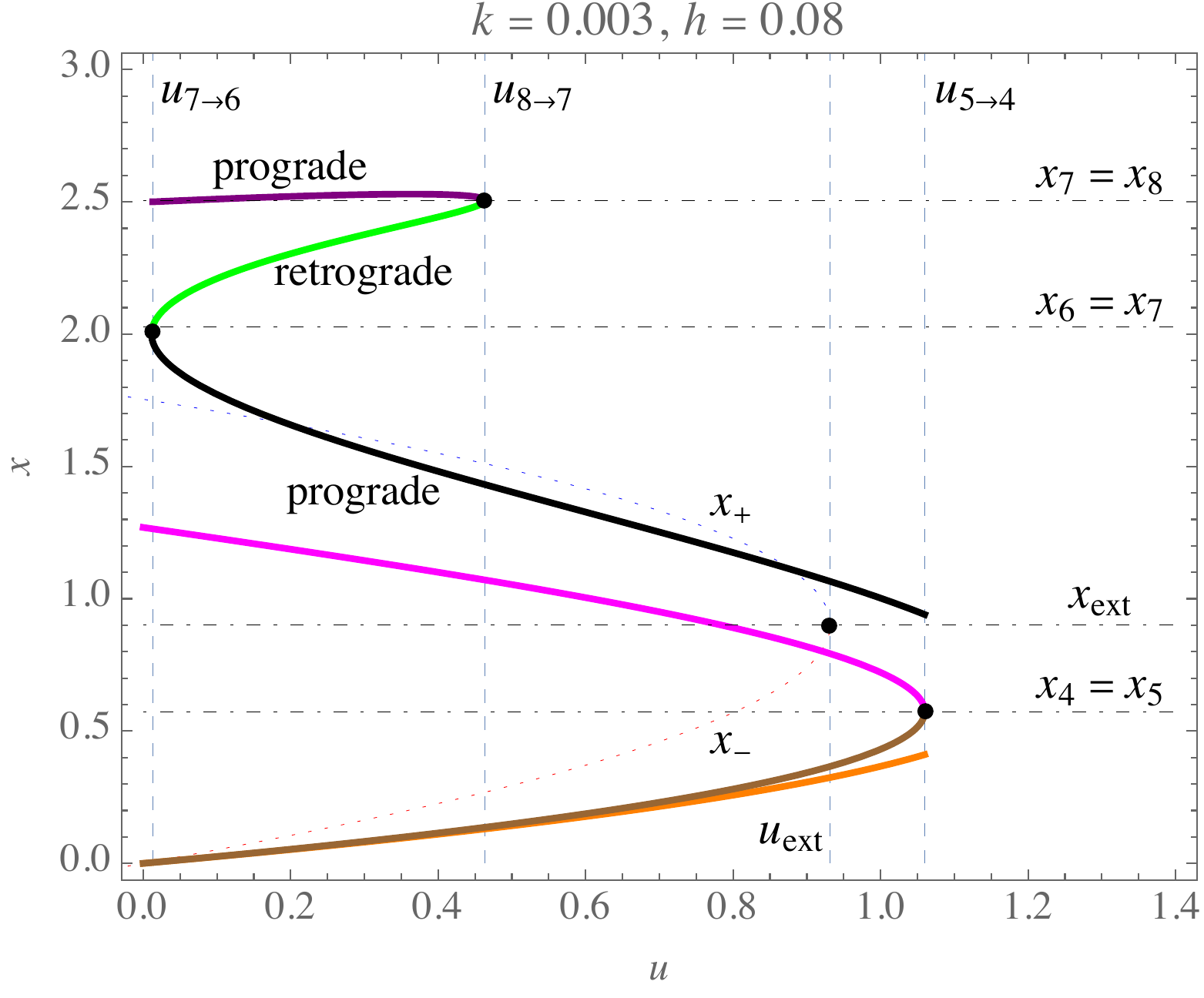}~(a)\qquad
     \includegraphics[width=8cm]{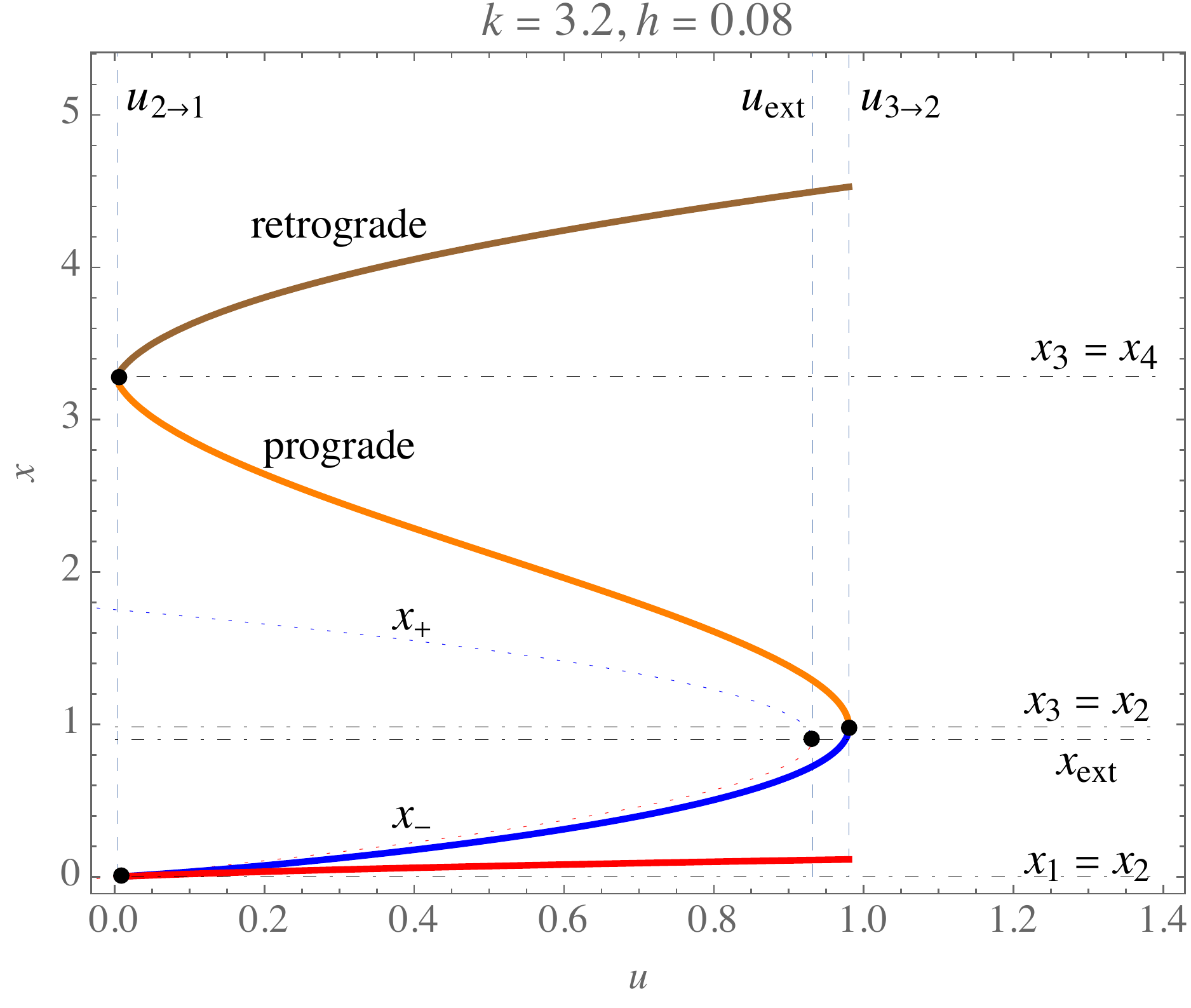}~(b)
    \caption{The $x-u$ diagrams for the real parts of the solutions $x_i$, plotted for $h=0.08$, $l=10^{-6}$, (a) $k=0.003$ and (b) $k=3.2$.}
    \label{fig:x-u0.003}
\end{figure}
We first consider $k_0=0.003$, which lies in the energy domain of Fig.~\ref{fig:x-k0.3}(b), and corresponds to the evolution of $x_7$ and $x_8$. The $u$-profiles of the solutions have been shown in the left panel of Fig. \ref{fig:x-u0.003}. Like before, we now let $u_{i\rightarrow j}$ to notate the values of the spin parameter, at which, $x_i=x_j$. As expected from the $x-k$ diagrams, $x_1$ and $x_2$ are totally complex-valued for this particular case, so they are absent from the profiles. Furthermore, the real parts of $x_8$ and $x_9$ coincide. In this diagram, we have also shown how the event and Cauchy horizons evolve until they reach the extremal radius. The solutions $x_3, x_4, x_5$ and $x_6$ are inside the event horizon for this energy, but they are real-valued up to $u=u_{5\rightarrow4}=1.06>u_{\mathrm{ext}}$, so that will be available, as well, around a naked singularity. At this point, the solutions $x_4$ and $x_5$ meet at $x=0.574$. The two other solutions $x_7$ and $x_8$ exist outside the event horizon, and by increasing the spin parameter up to $u_{8\rightarrow 7}=0.463$, they finally coincide at $x=2.505$. Note that, $x_7$ and $x_8$ correspond, respectively, to the \textit{retrograde} and \textit{prograde} planar orbits  for this choice of energy, and are confined in the range $1.972<x<2.505$ within the domain $u_{7\rightarrow6}=0.013<u<u_{8\rightarrow7}=0.463$. 

Let us now consider another example by choosing $k_0=3.2$, which corresponds to the unbound orbits in Fig.~\ref{fig:x-k0.3}(a). In this case, $x_6,x_7,x_8,x_9\in\Bbb{C}$, and the real-valued parts of $x_4$ and $x_5$ coincide. The solutions $x_1,x_2,x_3<x_+$ are extended beyond $u_{\mathrm{ext}}$, up to $u_{3\rightarrow2}=0.980$, at which $x_2=x_3=0.984$. Hence they are available around the naked singularity. The other two solutions $x_3,x_4>x_+$ coincide at $x=3.367$ for $u_{2\rightarrow1}=0.004$, the spin parameter value at which, it is also $x_1=x_2=0.001$. Note that, $x_3$ and $x_4$ correspond, respectively, to prograde and retrograde planar orbits. 

Now that we have established that bound orbits outside the event horizon can be present on $x_7$ and $x_8$ for the above examples, we can now inspect their stability. In fact, the stability condition for spherical particle orbits respects the additional condition $R''(x)\geq0$. From Eq.~\eqref{eq:R} and the changes of variables in Eqs. \eqref{eq:x,u,l,h}, one can infer that
\begin{equation}
R''(x)=30 l x^4-20 h x^3+12\Big[ k l u- (1-k)\Big]x^2+6 (2- h k u)x-2u (1-k).
    \label{eq:Rpp(x)}
\end{equation}
\begin{figure}[t]
    \centering
    \includegraphics[width=8cm]{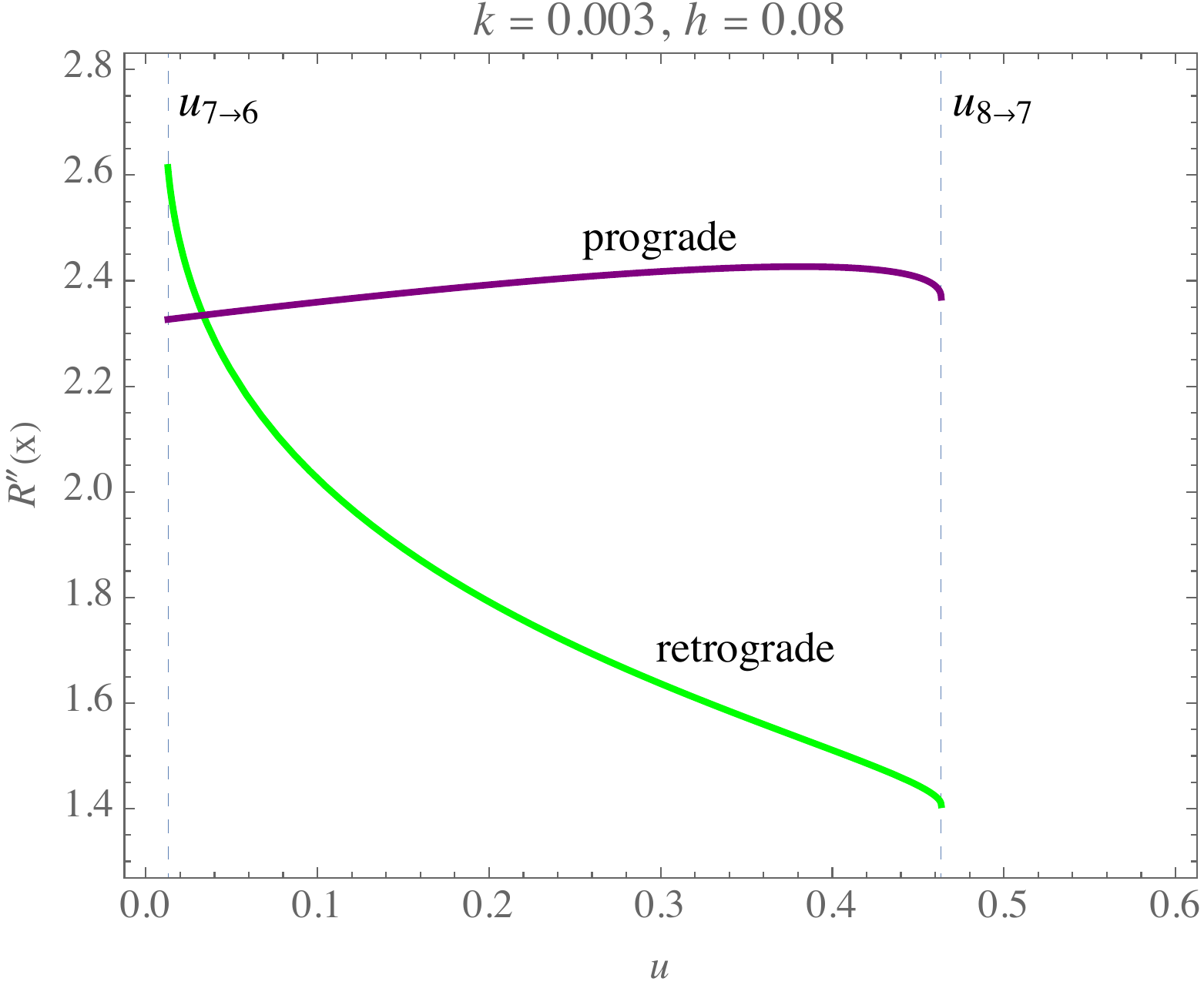}
    \caption{The $u$-profile of $R''(x)$ plotted for $x_7$ and $x_8$, in accordance with Fig.~\ref{fig:x-u0.003}(a).}
    \label{fig:Rpp-u0.003}
\end{figure}
In Fig.\ref{fig:Rpp-u0.003}, the $u$-profile of the above function is plotted for the two radii $x_7(u)$ and $x_8(u)$ in agreement with the parameter values considered in Fig.\ref{fig:x-u0.003}(a). As expected, the values of $R''(x)$ for these two solutions are restricted to the range $u_{7\rightarrow6}<u<u_{8\rightarrow7}$, and in both cases $R''(x)>0$. Hence, the values of $x_7$ and $x_8$ correspond to stable circular orbits located outside the event horizon. These orbits can, therefore, serve as generators of the ISCO for the RdRGT black hole when $k = 0.003$ and $u = 0.3$. On the other hand, to obtain the OSCO for this spin parameter, we can choose either $x_5$ or $x_6$, while ensuring that the energy range satisfies $k_{6\rightarrow5} = 0.894 \leq 1$.

\subsubsection{Non-monotonic behavior of the solutions}\label{subsubsec:nonmonotonic}

The above examples provide numerical solutions for determining the radii of prograde and retrograde circular orbits. Table \ref{tab:1} presents additional examples obtained by solving the nonic equation \eqref{eq:p9} for the cases of $k=0,1$, and $2$, in accordance with the parameter values shown in Figs. \ref{fig:x-k0.3}, \ref{fig:x-k0.9}, and \ref{fig:x-u0.003}. This table offers a clearer understanding of the variations in these radii outside the event horizon for the aforementioned cases. It should be noted that, under certain circumstances, prograde or retrograde trajectories with nonzero energies are only possible inside the event horizon, which is why they are not included in the table.
\begin{table}[t]
    \centering
    \begin{tabular}{c|c|c|c||c|c|c|c||c|c|c|c}
        $k$ & $u$ & $x_{\mathrm{prograde}}$ & $x_{\mathrm{retrograde}}$ & $k$ & $u$ & $x_{\mathrm{prograde}}$ & $x_{\mathrm{retrograde}}$ & $k$ & $u$ & $x_{\mathrm{prograde}}$ & $x_{\mathrm{retrograde}}$\\
        \hline\hline
        0 & 0.1 & 1.762 & 2.213, 2.517 & 1 & 0.1 & -- & 4.047 & 2 & 0.1 & 2.918 & --\\
        0 & 0.2 & 1.652 & 2.295, 2.533 & 1 & 0.2 & -- & 4.096 & 2 & 0.2 & 2.677 & --\\
        0 & 0.3 & 1.562 & 2.357, 2.549 & 1 & 0.3 & -- & 4.144 & 2 & 0.3 & 2.473 & --\\
         0 & 0.4 & 1.480 & 2.408, 2.564 & 1 & 0.4 & -- & 4.193 & 2 & 0.4 & 2.289 & --\\
          0 & 0.5 & 1.403 & 2.452, 2.578 & 1 & 0.5 & -- & 4.242 & 2 & 0.5 & 2.115 & --\\
          0 & 0.6 & 1.328 & 2.491, 2.593 &  1 & 0.6 & -- & 4.290 & 2 & 0.6 & 1.944 & --\\
           0 & 0.7 & 1.252 & 2.526, 2.606 & 1 & 0.7 & -- & 4.338 & 2 & 0.7 & 1.769 & --\\
            0 & 0.8 & 1.175 & 2.559, 2.619 & 1 & 0.8 & 1.730 & 2.269, 4.386 & 2 & 0.8 & 1.578 & --\\
             0 & 0.9 & 1.092 & 2.590, 2.632 & 1 & 0.9 & 1.335 & 2.437, 4.432 & 2 & 0.9 & 1.345 & --\\
              0 & 1.0 & 1.0 & 2. 618, 2.644 &  1 & 1.0 & -- & 2.529, 4.478 & 2 & 1.0 & -- & --
    \end{tabular}
    \caption{The radii of prograde and retrograde orbits outside the event horizon for different energies and spin parameters, obtained by assuming $h=0.08$ and $l=10^{-6}$.}
    \label{tab:1}
\end{table}
Here, we argue that these radii can also be calibrated by finding the critical values of the spacetime parameters. To fulfill this task, let us calculate the discriminant of the nonic equation \eqref{eq:p9} with respect to the $u$-parameter, which is obtained as
\begin{multline}
\delta_{\Delta_9} = h \bigg[-432 k^3 x^4+4 k^2 x \left(-72 l x^5+216 x^3-414 x^2\right)-4 k \left(-72 l x^6+186 l x^5+108 x^4-414 x^3+360 x^2\right)\bigg]\\
+144 k^4 x^3+48 k^3 x^2 \left(4 l x^3-9 x+15\right)+4 k^2 x \left(16 l^2 x^6-96 l x^4+180 l x^3+108 x^2-360 x+288\right)\\
-4 k \bigg[16 l^2 x^7-40 l^2 x^6-48 l x^5+180 l x^4-6 (26 l-6) x^3-180 x^2+288 x-144\bigg]+4 l^2 (5-2 x)^2 x^5,
    \label{eq:Deltap9u}
\end{multline}
up to first order of $h$, that provides the critical value 
\begin{multline}
h_{\mathrm{crit}} = \frac{1}{6 k x^2 \Big[12 (k-1) l x^4+18 (k-1)^2 x^2+69 (k-1) x+31 l x^3+60\Big]}\bigg[
36 k^4 x^3+12 k^3 x^2 \left(4 l x^3-9 x+15\right)\\
+4 k^2 x \left(4 l^2 x^6-24 l x^4+45 l x^3+27 x^2-90 x+72\right)-4 k \Big(4 l^2 x^7-10 l^2 x^6-12 l x^5+45 l x^4\\
+(9-39 l) x^3-45 x^2
+72 x-36\Big)+l^2 (5-2 x)^2 x^5
\bigg]
    \label{eq:hcritical}
\end{multline}
for which, $\delta_{\Delta_9} = 0$. By means of this critical value, one can determine the turning points of the polynomial \eqref{eq:p9}. For the zero-energy particles with $k=0$, we get
\begin{equation}
h_{\mathrm{crit}}^{0} = \frac{2 \Big[4 l^2 x^7-10 l^2 x^6-12 l x^5+45 l x^4+(9-39 l) x^3-45 x^2+72 x-36\Big]}{3 x^2 \left(12 l x^4-31 l x^3-18 x^2+69 x-60\right)}.
    \label{eq:hcritk0}
\end{equation}
Furthermore, for $k=1$ and $k=2$ we obtain
\begin{eqnarray}
&& h_{\mathrm{crit}}^{1}=\frac{l^2 (2 x+5)^2 x^5+156 l x^3+144}{6 x^2 \left(31 l x^3+60\right)},\label{eq:hcritk1}\\
&& h_{\mathrm{crit}}^{2} = \frac{36 l^2 x^7+60 l^2 x^6+l (25 l+96) x^5+360 l x^4+24 (13 l+3) x^3+360 x^2+576 x+288}{12 x^2 \left(12 l x^4+31 l x^3+18 x^2+69 x+60\right)}.\label{eq:hcritk2}
\end{eqnarray}
Now to find the desired radii, we first consider examples of $h_0$ and equal them to $h_{\mathrm{crit}}^{i} \,\, (i=0,1,2)$. Using the Eqs. \eqref{eq:hcritk0}--\eqref{eq:hcritk2}, the values of $x_0$ can be obtained for each of the energy cases $k=0,1$ and $2$. The nonic equation $p_9(x_0,u,\left\{k_0=0,1,2\right\},h_0,l_0)=0$, is then solved for $u$, to find the corresponding $u_0$ for each of the cases. This spin parameter is again substituted in the nonic equation $p_9(x,u_0,\left\{k_0=0,1,2\right\},h_0,l_0)=0$, to find the $x$-solutions. For example, for $k_0=0$, $h_0=0.08$ and $l_0=10^{-6}$, the prograde and retrograde radii are obtained as $x_{\mathrm{prograde}}=1.542$ and $x_{\mathrm{retrograde}} = 2.367$, corresponding to $u_0 = 0.319$, which are in conformity with the data given in Table \ref{tab:1}.

In this section, we have examined the characteristic equations governing the radii of planar orbits around the RdRGT black hole. These orbits correspond to the case of $\nu=0$ in the main equation \eqref{eq:p14}. For the investigation of polar orbits, characterized by $\nu=1$ in Eq. \eqref{eq:p14}, once again a polynomial equation of fourteenth order is obtained. However, instead of directly analyzing this equation, we can exploit the fact that for polar orbits, $L_c=0$. By utilizing Eq. \eqref{eq:L_c} and performing some algebraic manipulations, we arrive at the octic equation
\begin{multline}
4 l x^8
- \left(3 h+14 l\right) x^7
+  \left(11 h+4 k+6 l u+10 l-2\right)x^6\\
- \left(5 h u+8 h+8 k+6 l u-10\right) x^5 
+ \left(5 h u+4 k u+6 k-8 u-16\right)x^4\\
+ 4\left(2-2 k u+3 u\right) x^3 
+ 2\left( k u^2+2 k u- u^2-4 u\right)x^2
+2 u^2 x
-2 k u^2=0.
    \label{eq:p8}
\end{multline}
{Indeed, the methods employed to analyze the behavior of the solutions of the aforementioned equation are quite similar to those discussed earlier for the nonic equation \eqref{eq:p9}. Consequently, we can close this section at this point and proceed with our study by calculating the analytical solutions for the orbits. With these solutions available, it becomes feasible to perform simulations of spherical orbits around the RdRGT black hole.}

{However, before proceeding to the next section, it is crucial to highlight some important features of stable circular orbits. In particular, solutions such as $x_7$ and $x_8$ correspond to extremums in the radial effective potential experienced by particles as they approach the black hole. These radii give rise to circular orbits within the equatorial plane. 
When these extremums represent minima in the effective potential (as in the cases of $x_7$ and $x_8$), the stability of these orbits implies that deviations from these minima result in unstable orbits. In such cases, particles exhibit Keplerian motion, oscillating between two points known as the periapsis and apoapsis of their trajectories. It is important to note that, unlike circular orbits, these Keplerian trajectories are non-planar and correspond to planetary bound orbits.
However, when deviations from the potential minima become significant and the turning points approach the maximum of the effective potential, marginally bound orbits occur. At this point, particles have the opportunity to escape the black hole. The maximum of the effective potential exhibits a certain degree of stability, allowing for the existence of spherical orbits. Nevertheless, these orbits are highly sensitive to radial perturbations, which can cause the transition from bound to unbound motion. Hence, particles can deflect and escape from the black hole when approaching the maximum of the effective potential.
In summary, the bound and marginally bound circular orbits encompass two additional motion possibilities; particles can either remain confined to the black hole or escape from its gravitational field.}

{Now that we have discussed the classifications of particle motion in black hole spacetimes, we can now continue with the derivation of the analytical solutions for the equations of motion governing spherical orbits.}



\section{Analytical solutions for the spherical particle orbits}\label{sec:AnlyticalSol}

In this section, we derive the analytical solutions for the evolution equations governing the polar and azimuth angles of spherical orbits, as described in Section \ref{sec:geodesic}. As discussed previously, the classification of geodesics is based on the energy values, which determine whether the particle motion is bound or unbound. The solutions encompass both planar and non-planar orbits, corresponding to the cases of $\nu=0$ and $\nu\neq 0$ respectively.


\subsection{The latitudinal motion}\label{subsec:thetaMotion}

Considering the definitions in Eqs.~\eqref{eq:x,u,l,h} and \eqref{eq:Q_new}, one can re-write the evolution equation \eqref{eq:dthetadtau} as
\begin{equation}
-\frac{\ed Z}{\ed\lambda} = \sqrt{\Theta_Z},
    \label{eq:dZdlambda}
\end{equation}
in which $Z=\cos\theta$, and for bound orbits of massive particles (i.e. $k\leq1$), we specify 
\begin{equation}
\Theta_Z =\left(1-k\right)u Z^4-\left[\left(1-k\right)u-\frac{L^2}{1-\nu}\right]Z^2+\frac{L^2\nu}{1-\nu}.
    \label{eq:Theta_Z}
\end{equation}
Naturally, for planar orbits with $\nu=0$, the above expression reduces to $\Theta_Z = (1-k)u Z^4-(L^2+u-uk)Z^2$, whereas for polar orbits with $L=0$, it becomes $\Theta_Z=(1-k)u Z^2(Z^2-1)$. Note that, Eq. \eqref{eq:Theta_Z} can help us determine the maximum and minimum latitudes reachable by particles, which are the angular values where $\Theta_Z=0$. This equation provides the values
\begin{equation}
 Z_{\substack{{\max}\\{\min}}}^2 = \frac{1}{2Z_0}\bigg[L^2+Z_0\pm\sqrt{\left(L^2+Z_0\right)^2-4L^2\nu Z_0}
\bigg],\label{eq:Zmax,Zmin}
\end{equation}
with $Z_0 = (1-k)(1-\nu)u$, that confine the $Z$-parameter. In this way, the polar angle $\theta$ oscillates in the domain $\theta\in[\theta_{\min},\theta_{\max}]$, in which $\theta_{\min}=\arccos(Z_{\max})$ and $\theta_{\max} = \arccos(Z_{\min})$. Note that, there is a sign change for the azimuth angle $\phi$, where ${\ed\phi}/{\ed\lambda}=0$, and can be determined by means of Eq.~\eqref{eq:dphidtau}. After doing the substitutions, this angular turning point is obtained as
\begin{equation}
Z_t^2 = \frac{\Big(\sqrt{k u}-L\Big)\Big(lx^4-hx^3+2x\Big)+Lx^2}{\sqrt{ku}\Big(lx^4-hx^3+2x\Big)-lu}.
    \label{eq:Zt}
\end{equation}
Hence, one can infer that the physically acceptable polar segments are where $|Z_t|<|Z_{\max}|$. Now, to obtain the exact analytical solution to the evolution of the polar angle, we directly integrate Eq.~\eqref{eq:dZdlambda}, which after proper substitutions results in
\begin{equation}
\theta(\lambda) = \arccos\left(Z_{\max}-\frac{3}{\psi_2-12\wp(\sqrt{\psi}\,\lambda)}\right),
    \label{eq:theta_lambda}
\end{equation}
in which $\wp(\cdots)\equiv\wp(\cdots;\bar{g}_2,\bar{g}_3)$, is the Weierstrassian $\wp$ elliptic function with the invariants
\begin{subequations}
\begin{align}
    & \bar{g}_2 = -\frac{1}{4}\left(\psi_1-\frac{\psi_2^2}{3}\right),\\
    & \bar{g}_3 = -\frac{1}{16}\left(\frac{2\psi_2^3}{27}-\frac{\psi_1\psi_2}{3}-\psi_0\right),
\end{align}
\label{eq:barg2g3}
\end{subequations}
where
\begin{subequations}
\begin{align}
  & \psi_0 = \frac{(1-k) u }{\psi},\\
  & \psi_1 = \frac{4 Z_{\max} (1-k) u}{\psi},\\
  & \psi_2 = -\frac{1}{\psi}\left[
  (1-k)u-\frac{L^2}{1-\nu}-6Z_{\max}^2(1-k)u
  \right],\\
  & \psi =Z_{\max}\left[(1-k) u Z_{\max}^2+(1-k) u-\frac{L^2}{1-\nu }\right]-Z_{\max} \left[\frac{L^2}{1-\nu }-(1-k) u\right]+3 (1-k) u Z_{\max}^3.
\end{align}
\label{eq:psi012}
\end{subequations}
The above solution can also be used to determine the Mino time at which the $\theta$-profile periodically traverses the equatorial plane and generates \textit{nodes}. Thus, for each full oscillation of the $\theta$-coordinate, the profile encounters two nodes. According to Eq.~\eqref{eq:theta_lambda} we get
\begin{equation}
\lambda_{\mathrm{nod}} = \frac{1}{\sqrt{\psi}}\wp^{-1}\left(
\frac{\psi_2}{12}-\frac{1}{4 Z_{\max}}
\right),
    \label{eq:lambda_node}
\end{equation}
for $\theta=\pi/2$, where $\wp^{-1}(\cdots)\equiv\wp^{-1}(\cdots;\bar{g}_2,\bar{g}_3)$.
Note that, for unbound orbits with $k>1$, the form of the analytical solution \eqref{eq:theta_lambda} remains the same, however, the Weierstrass invariants \eqref{eq:barg2g3} are affected through the exchange $(1-k)\rightarrow(k-1)$ in the quantities in Eqs.~\eqref{eq:psi012}.

\subsection{The azimuth motion}\label{subsec:phiMotion}

The solution to the evolution equation \eqref{eq:dphidtau} can be written as
\begin{equation}
\phi(\lambda) = \mathcal{C}(x_i)\Phi_{\theta_1}(\lambda)+\Phi_{\theta_2}(\lambda),
    \label{eq:Phi(lambda)_0}
\end{equation}
for spherical particle orbits on the radii $x_i$, where
\begin{subequations}\label{eq:cpPhitheta}
\begin{align}
    & {\mathcal{C}}(x_i)=\frac{\sqrt{u k}
    }{\Delta(x_i)}\Big[
   x_i^2+u-u\,\xi_c(x_i)-\Delta(x_i)\Big],\label{eq:ci}\\
    & \Phi_{\theta_1}(\lambda) = \int^{\theta(\lambda)}\frac{\ed\theta}{\sqrt{\Theta(\theta)}},\label{eq:Phitheta1}\\
    & \Phi_{\theta_2}(\lambda) = \xi_c\int^{\theta(\lambda)}\frac{\ed\theta}{\sin^2\theta\sqrt{\Theta(\theta)}},\label{eq:Phitheta2}
\end{align}
\end{subequations}
in which $\xi_c(x)=L_c(x)/\sqrt{k}$, is the impact parameter of particles. The integral \eqref{eq:Phitheta1} can be directly calculated from the solution \eqref{eq:theta_lambda}, which provides
\begin{equation}
\Phi_{\theta_1}(\lambda)=\frac{1}{\sqrt{\psi}}\wp^{-1}\left(
\frac{\psi_2}{4}-\mathcal{U}_\theta
\right),
    \label{eq:Phitheta1_sol}
\end{equation}
where
\begin{equation}
\mathcal{U}_\theta=\frac{\psi_2}{6}+\frac{1}{4\left[Z_{\max}-\cos\theta\right]}.
    \label{eq:utheta}
\end{equation}
In addition, the integral \eqref{eq:Phitheta2} can be solved by the same methods, yielding the following:
\begin{equation}
\Phi_{\theta_2}(\lambda) = \mathcal{K}_0\left[
\mathcal{K}_1\mathscr{F}_1(\mathcal{U}_\theta)-\mathcal{K}_2\mathscr{F}_2(\mathcal{U}_\theta)-\sqrt{\psi}\lambda
\right],
    \label{eq:Phitheta2_sol}
\end{equation}
where
\begin{subequations}\label{eq:K0,1,2}
\begin{align}
    & \mathcal{K}_0 = \frac{\xi_c^2}{
    u(1-Z_{\max})(1+Z_{\max})\sqrt{2Z_{\max}\left(Z_{\max}^2+Z_0^2\right)}},\label{eq:K0,1,2a}\\
    & \mathcal{K}_1 = \frac{1+Z_{\max}}{8(1-Z_{\max})},\label{eq:K0,1,2b}\\
     & \mathcal{K}_2 = \frac{1-Z_{\max}}{8(1+Z_{\max})},\label{eq:K0,1,2b}
\end{align}
\end{subequations}
and 
\begin{equation}
\mathscr{F}_j(\mathcal{U}_\theta) = \frac{1}{\wp'(\upsilon_j)}\left[
\ln\left(
\frac{\sigma\left(\ss(\mathcal{U}_\theta)-\upsilon_j\right)}{\sigma\left(\ss(\mathcal{U}_\theta)+\upsilon_j\right)}
\right)+2\ss(\mathcal{U}_\theta)\zeta(\upsilon_j)
\right],\qquad j=1,2,
\label{eq:Fj}
\end{equation}
in which $\wp'(\upsilon) = \ed\wp(\upsilon)/\ed \upsilon$. Furthermore, $\sigma(\cdots)$ and $\zeta(\cdots)$ are the Sigma and Zeta Weierstrassian elliptic functions, and the Weierstrass invariants are the same as those given in Eq. \eqref{eq:barg2g3}. In Eq. \eqref{eq:Fj}, we have defined
\begin{subequations}\label{eq:upsilonUtheta}
\begin{align}
    & \upsilon_1=\wp^{-1}\left(\frac{\psi_2}{12}-\frac{1}{4|1-Z_{\max}|}\right),\label{eq:upsilon1}\\
    & \upsilon_2=\wp^{-1}\left(\frac{\psi_2}{12}-\frac{1}{4|1+Z_{\max}|}\right).\label{eq:upsilon2}
\end{align}
\end{subequations}
In order to simulate the orbits, it is convenient choosing $\phi(Z_{\max})=0$. Then by interpolating $\theta\rightarrow\theta(\lambda)$ in the expression \eqref{eq:Phitheta1_sol}, the evolution of the $\phi$-coordinate is obtained explicitly.

\subsection{Explicit examples of non-planar orbits}\label{subsec:examples}

\begin{figure}[t]
    \centering
    \includegraphics[width=4cm]{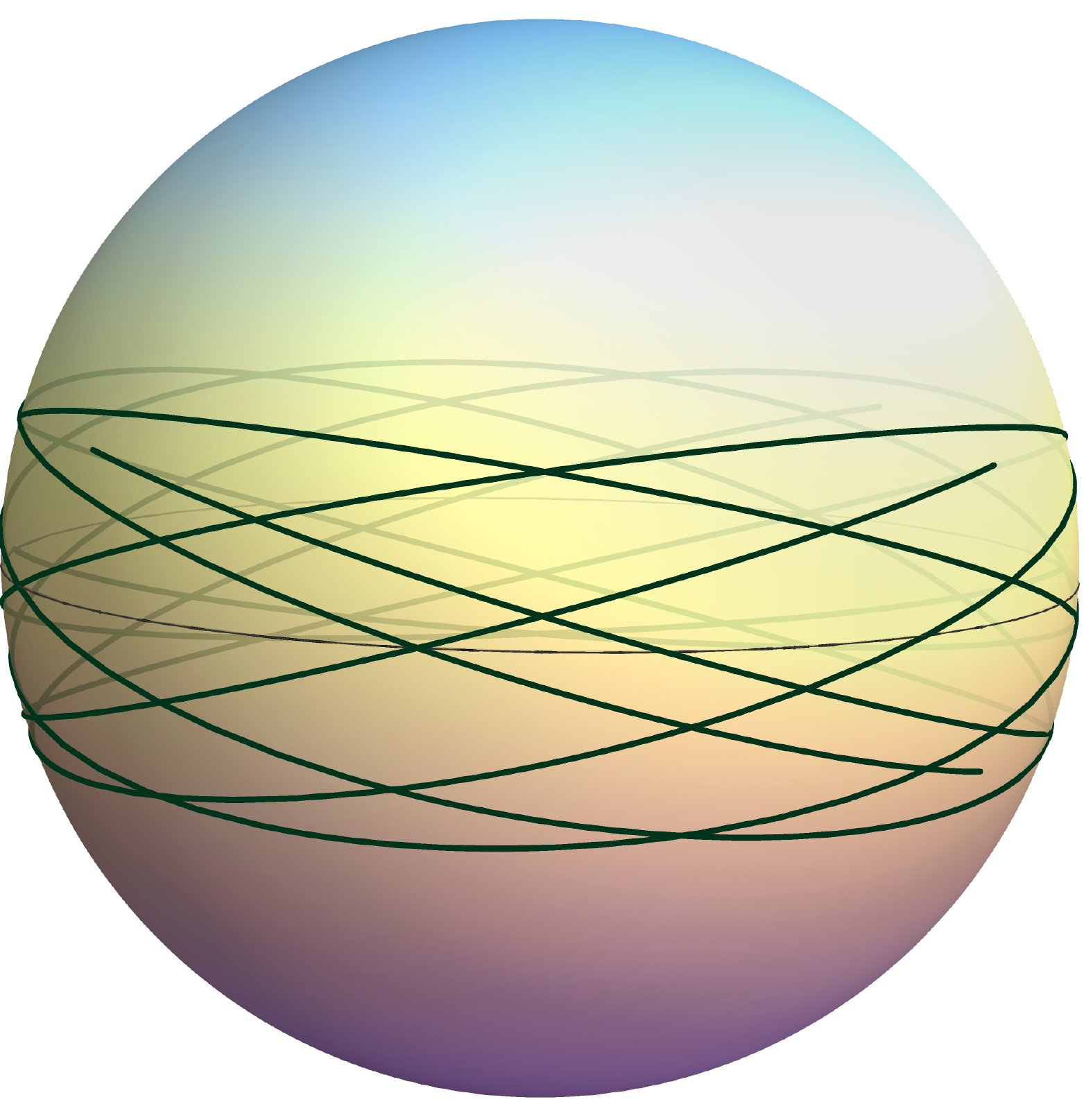}~(a)
     \includegraphics[width=4cm]{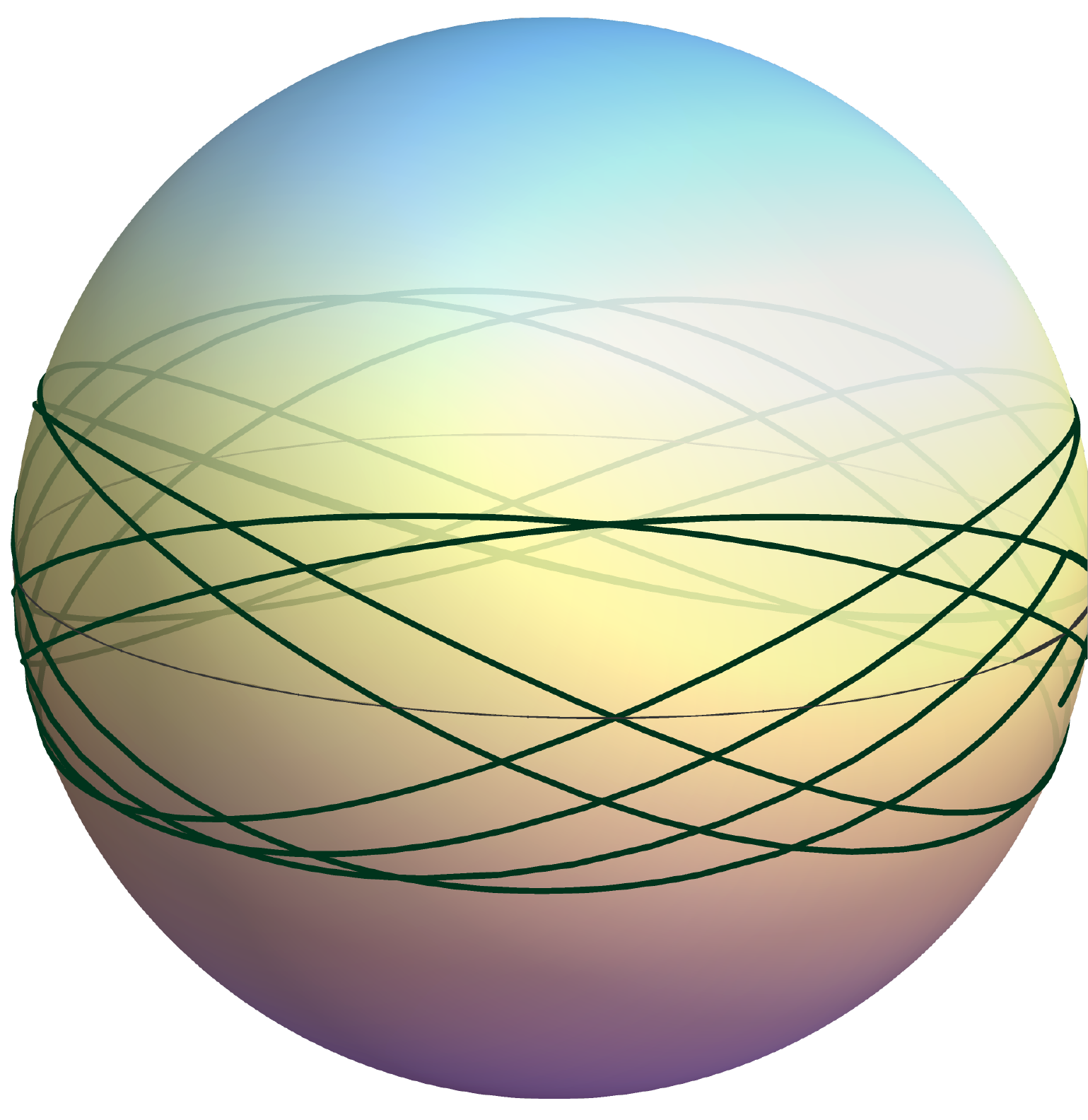}~(b)
      \includegraphics[width=4cm]{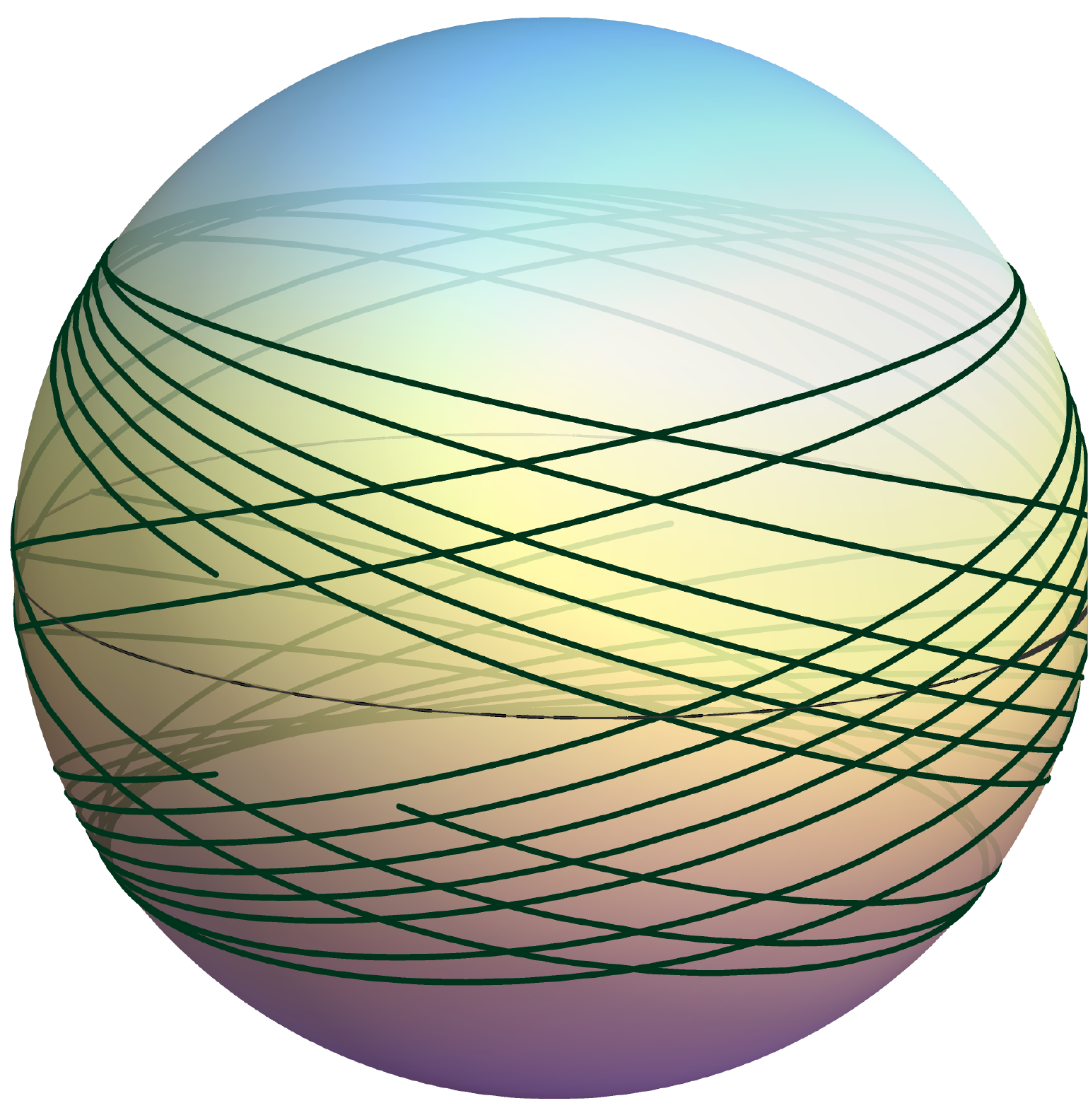}~(c)
       \includegraphics[width=4cm]{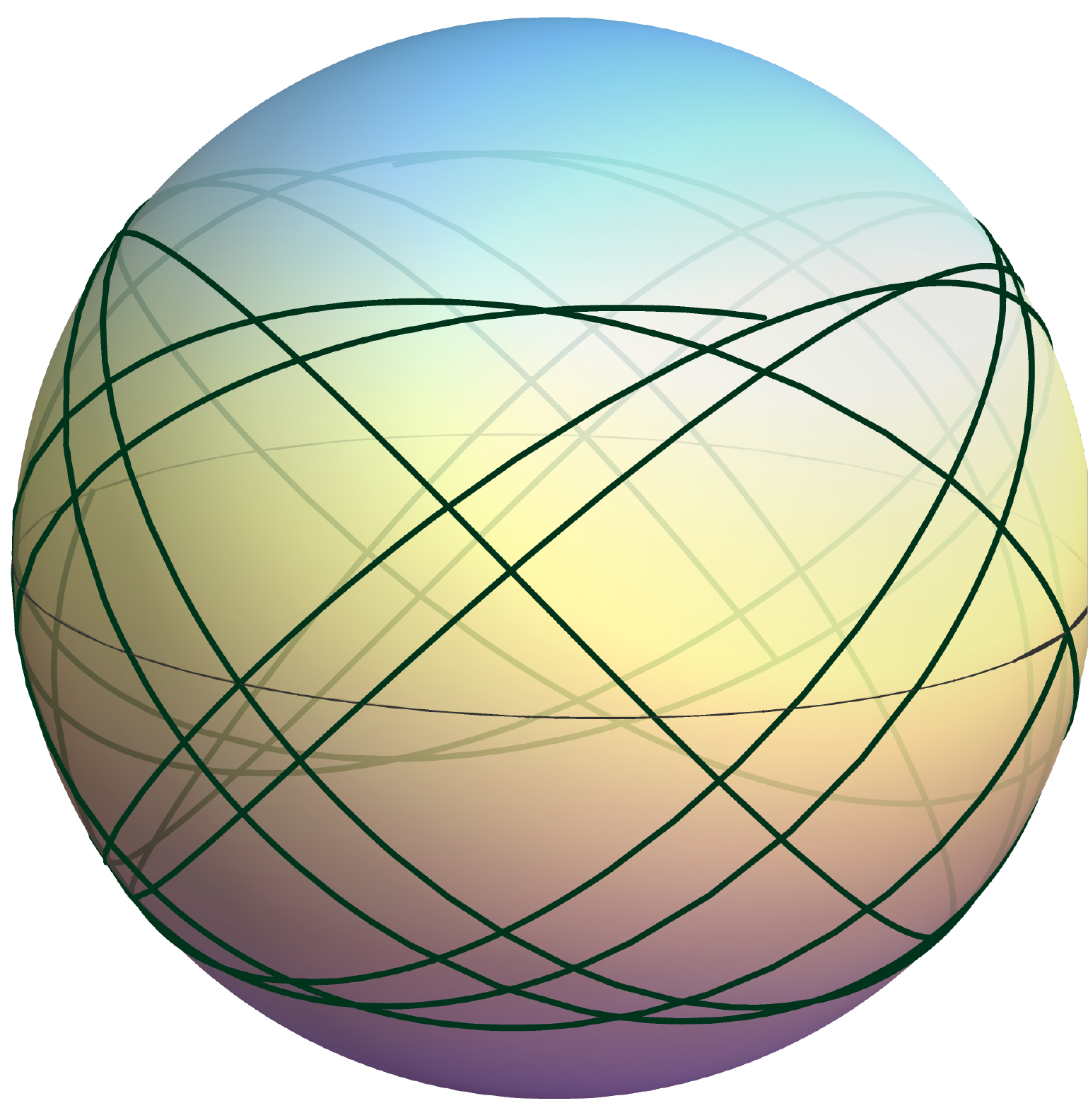}~(d)
        \includegraphics[width=4cm]{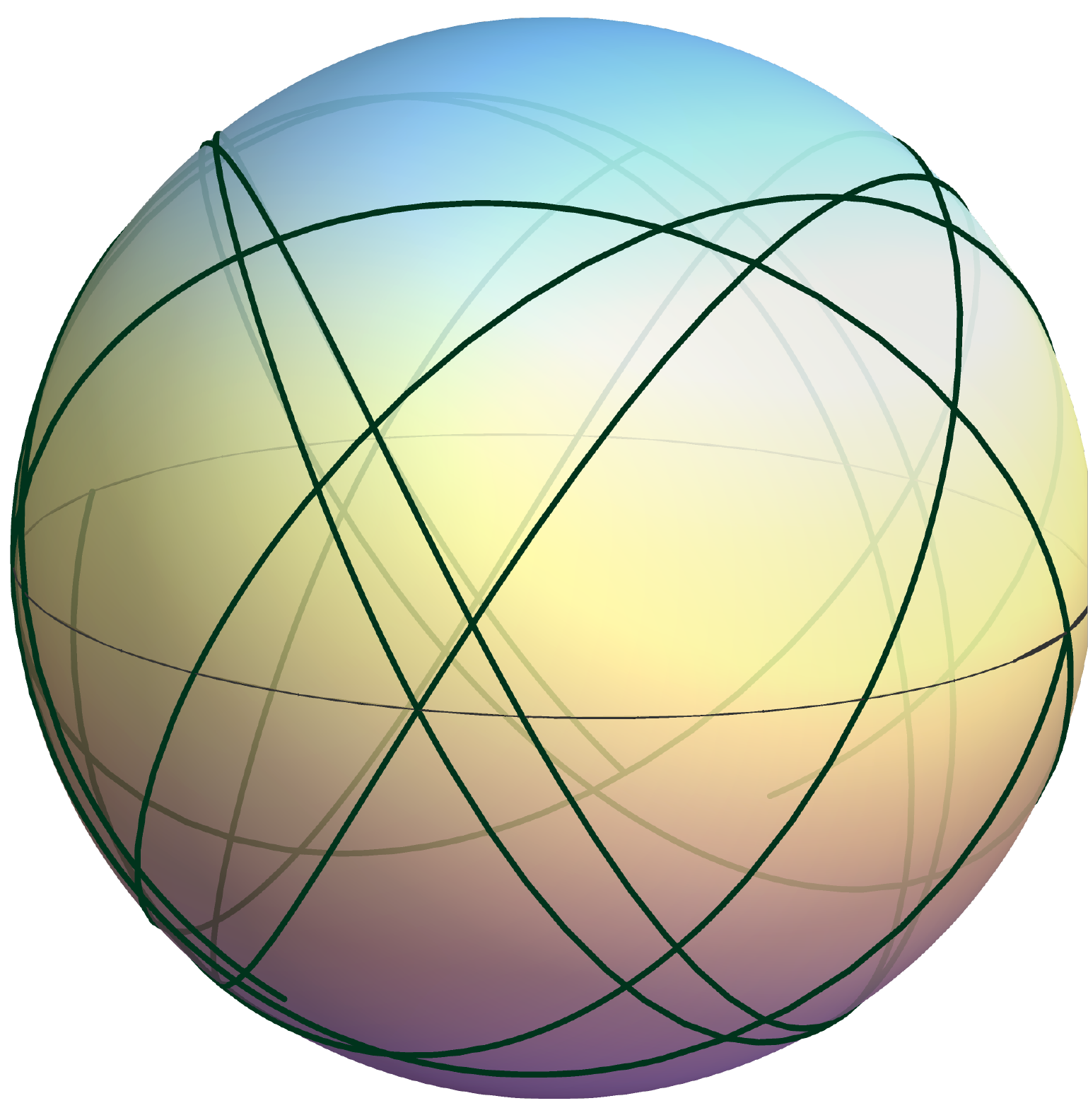}~(e)
         \includegraphics[width=4cm]{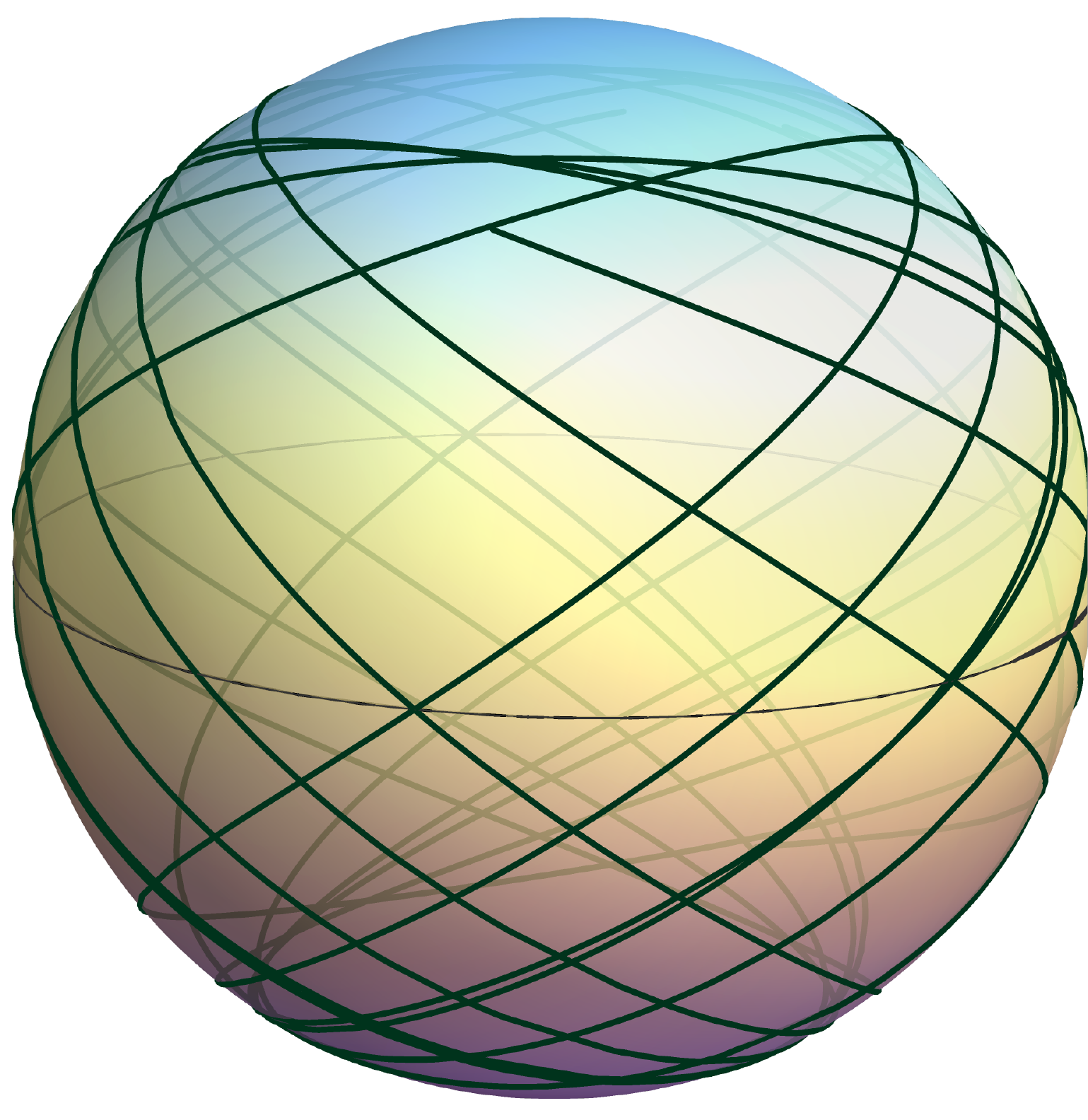}~(f)
          \includegraphics[width=4cm]{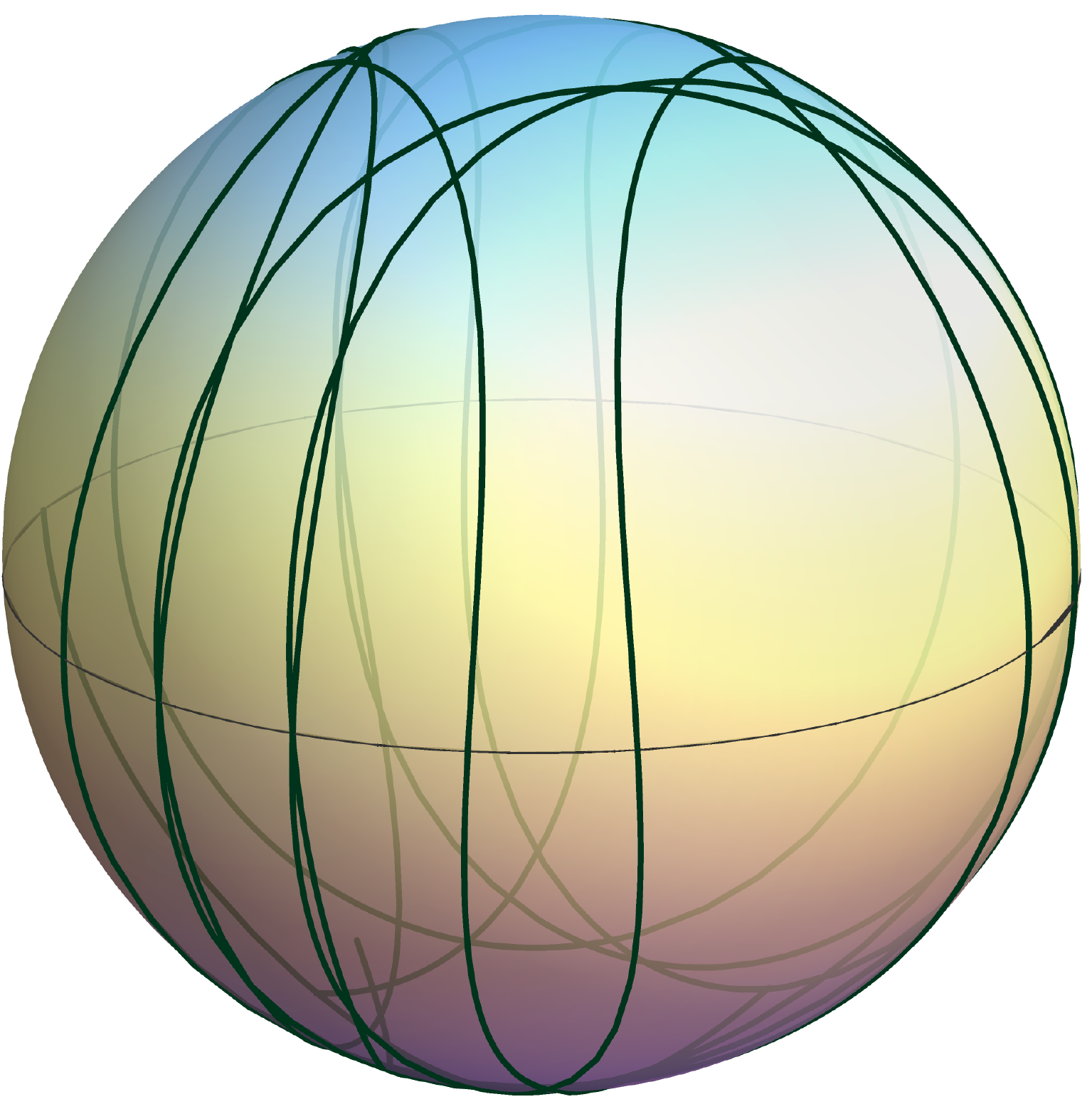}~(g)
           \includegraphics[width=4cm]{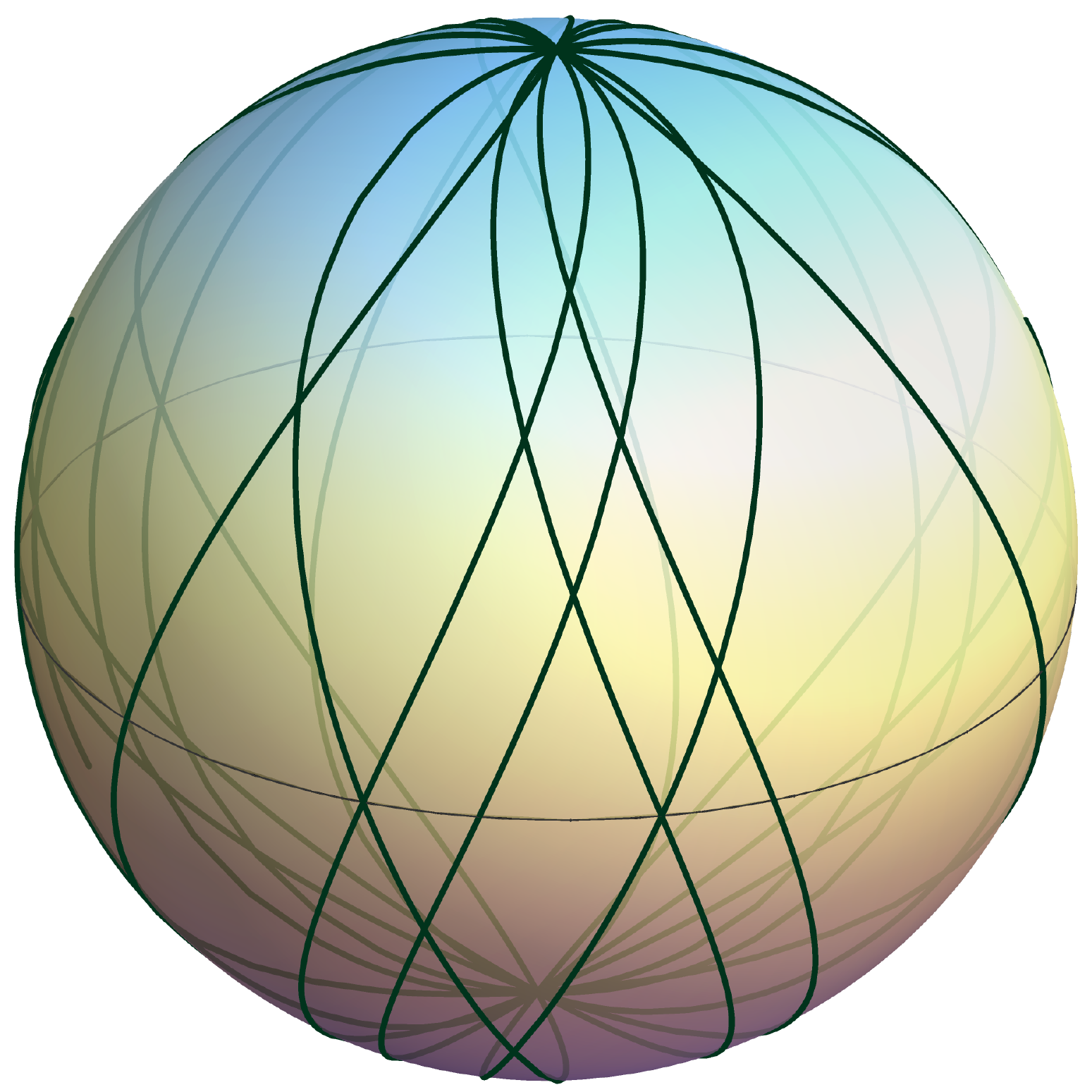}~(h)
    \caption{Some examples of spherical particle orbits in accordance with the data presented in Table \ref{table:2}. The sphere indicates the closure of points swept by the radii $x_i$, which is cut to halves by a circle on the $\theta=\pi/2$ surface.}
    \label{fig:orbitsGeneral}
\end{figure}

We can now apply the analytical solutions obtained above, to make some illustrative examples of spherical particle orbits around the RdRGT black hole. To proceed with this, we first have to consider some specific initial conditions for the black hole parameters. We then span the analytical solutions \eqref{eq:theta_lambda} and \eqref{eq:Phi(lambda)_0} onto a Cartesian system
defined as
\begin{subequations}\label{eq:Kerr-Schild}
\begin{align}
    & \mathcal{X}=
    {x}\sin\theta\cos\phi,\label{eq:Kerr-Schilda}\\
    & \mathcal{Y}=
    {x}\sin\theta\sin\phi,\label{eq:Kerr-Schildb}\\
    & \mathcal{Z}=
    x\cos\theta.\label{eq:Kerr-Schildc}
\end{align}
\end{subequations}
In Fig.~\ref{fig:orbitsGeneral}, we have used the data given in Table \ref{table:2} to simulate some exemplary bound and unbound spherical orbits on slow and fast-rotating RdRGT black holes. The radii of orbits have been obtained by numerically solving the general equation \eqref{eq:p14}, for different inclinations.  
\begin{table}[t]
    \centering
    \begin{tabular}{c||c|c|c|c|c|c}
    name & $u$ & $\nu$ & $h$ & $k$ & $x_i$ & $\xi_c(x_i)$\\
    \hline\hline
        (a) & 0.3 & 0.1 & 0.08 & 0.003 & 2.342 & 7.093 \\
       (b)  &  0.9 & 0.1  & 0.08 & 0.890 & 1.921 & 2.208\\
       (c) & 0.85 & 0.3 & 0.04 & 0.91 & 5.038 & 6.590\\
       (d) & 0.5 & 0.4 & 0.004 & 1.1 & 2.533 & 3.938\\
       (e) & 0.6 & 0.6 & 0.0004 & 0.93 & 5.803 & 7.916 \\
       (f) & 0.8 & 0.7 & 0.01 & 1.3 & 4.109 & 2.711\\
       (g) & $u_{\mathrm{ext}}=0.990$ & 0.9 & 0.001 & 0.9 & 2.922 & 1.067\\
       (h) & 0.85 & 1 (polar) & 0.0001 & 1 & 3.497 & 0.0045 
    \end{tabular}
    \caption{The information for the exemplary cases outside the event horizon, considered for $l=10^{-6}$.}
    \label{table:2}
\end{table}
Referring to the table, case (a) is related to the solution $x_7$ in Fig.~\ref{fig:x-k0.3}(b), but here corresponds to a retrograde bound orbit above the equatorial plane. In the same sense, case (b) is related to the solution $x_6$, which for the selected energy value corresponds to a bound orbit above the equatorial plane. The other non-planar orbits are assumed as follows: Case (c) considers a retrograde bound orbit for a fast-rotating black hole. In case (d), a prograde unbound orbit is assumed. Case (e) once again assumes a retrograde bound orbit, which is here derived for low energies. Case (f) corresponds to a retrograde unbound orbit for a fast-rotating black hole. In case (g), we present an example of prograde bound orbits around an extremal black hole, and finally, in case (h), a polar bound orbit around a fast-rotating black hole is considered. As expected, increasing the inclination leads to orbits that span broader regions of the spherical surface, accompanied by an increase in the frequency of polar angle oscillations. This behavior persists until the orbits become entirely polar when $\nu=1$. Additionally, we observe that for the same inclination, an increase in the spin parameter reduces the difference between consecutive values of $\lambda_{\mathrm{nod}}$, as the orbits more frequently intersect the equatorial plane. Furthermore, we discover that a lower massive gravity parameter results in more uniform orbits.

Given that the solutions are fully analytic, these aforementioned findings can be replicated in different examples, allowing for simulations of various types of orbits. Consequently, we leave the discussion at this point and proceed to summarize our results in the following section.

\section{Summary and conclusions}\label{sec:conclusion}

In this paper, we presented a comprehensive analysis of the spherical orbits of massive particles around an RdRGT black hole. Our investigation encompassed both analytical and numerical approaches, with a focus on providing exact analytical solutions whenever feasible. Commencing our investigation, we provided a concise overview of the dRGT massive theory of gravity and its static black hole solution. Employing a modified version of the Newman-Janis algorithm, we proceeded to construct a rotating analog of the spacetime, specifically termed as the RdRGT solution. We then examined the causal structure of this spacetime by studying the fourth-order polynomial equation $\Delta(r) = 0$. We found that, under certain conditions, the equation's discriminant is positive, implying the presence of three horizons for the black hole. By considering particular parameter cases, we were able to determine specific ranges for the massive gravity parameter for the RdRGT black hole with three horizons. We obtained the analytical expressions for the radial positions of the horizons and complemented our findings with a numerical investigation. This numerical study aimed to illustrate the radial evolution of the black hole horizons and the static limit hypersurfaces for various values of the massive gravity parameter. We continued our investigation by presenting the nonlinear differential equations of motion for massive particles and derived the conditions for the motion invariants, that lead to spherical orbits. Subsequently, we derived the complete characteristic polynomial equation governing the radii of spherical particle orbits in the RdRGT spacetime. This equation is of the fourteenth order and encompasses all possible inclinations. To focus our analysis, we specifically examined planar orbits confined to the equatorial plane, resulting in a nonic equation. Numerical analysis of the solutions to this nonic equation was conducted, considering various aspects such as their evolution in terms of energy and spin parameter for specific cases. We observed certain discontinuities in the energy profiles; however, we found that the bound and unbound solutions remain connected under specific criteria. By calculating the connecting energies, we argued that there exist potentially bound and stable solutions beyond the event horizon. Additionally, we conducted a detailed numerical examination of the solutions' evolution in relation to variations in the spin parameter, distinguishing between prograde and retrograde particle orbits. Interestingly, we discovered that the solutions remain valid even in the presence of naked singularities. In all cases, we numerically calculated the intersection points for the energy and spin parameter, accompanied by their corresponding radial distances. Furthermore, through a comprehensive stability analysis, we established the existence of at least two stable solutions outside the event horizon, which could potentially serve as candidates for the ISCO around the RdRGT black hole. We terminated our study on the radii of spherical orbits by examining the critical values of the massive gravity parameter, at which, the nonic equation becomes degenerate. We demonstrated that by utilizing these critical values, we can obtain the solutions, effectively by regenerating the desired spin parameters associated with each of the roots. The obtained results were consistent with the previously determined radii of prograde and retrograde orbits for each specific case. In the final section of the paper, we focused on obtaining analytical solutions for the angular equations of motion. This involved solving the evolution equations for both the polar and azimuth angles in their most general forms. By directly integrating the original differential equations, we expressed the solutions in terms of the Weierstrassian elliptic functions, which covered both the bound and unbound orbits at constant radii. To further explore their practical implications, we considered specific cases involving definite values for the black hole parameters and the initial energy of the test particles. Utilizing the derived analytical solutions, we simulated the motion of the test particles on spherical orbits. This approach includes a gradual increase in the inclination while the overall massive gravity parameter decreases. We considered different spin parameters and energies to encompass both bound and unbound orbits around regular and extremal states of the RdRGT black hole. Our investigations revealed a variety of behaviors, ranging from nearly planar orbits to polar orbits. In this paper, we conducted numerical investigations on the solutions of a nonic equation and performed analytical studies on the equations of motion. Since the black hole spacetime of the RdRGT black hole extends that of the Kerr black hole, the results and methods presented in this study can be applied to different types of stationary spacetimes derived from other extended theories of gravity that involve different matter properties. 

{For example, one intriguing aspect that can be studied is the redshift and blueshift of particles trapped by the gravitational pull of the RdRGT black hole, using a similar approach as in Ref. \cite{kraniotis_gravitational_2021}. Furthermore, it is possible to investigate the shape of RdRGT black holes by examining the distribution of accreting material along orbits of constant radius. This can lead to predictions regarding the observed shapes of astrophysical black holes, such as M87* and Sgr A* \cite{universe6090154}. The methods used in this paper can also be applied to study the evolution of the ISCOs in dynamical spacetimes \cite{Song:2021}. Additionally, the reflection symmetry of the RdRGT black hole can be explored, which may provide insights into the existence of curved accretion disks \cite{chen_curved_2022}. Moreover, numerical analyses of the radii of planar orbits can be conducted for various axially symmetric spacetimes, serving as tests for the analytical solutions \cite{ospino_all_2022}. These methods can also be extended to investigate spherical particle orbits and accretion in stationary black holes with quintessential parameters or minimally coupled scalar fields \cite{PhysRevD.105.124039,PhysRevD.106.024034}. In addition, the study of gravitomagnetic phenomena, such as the Lense-Thirring effect, can be carried out for orbits of constant radius in stationary spacetimes \cite{baines_constant-r_2022}. All of these investigations can be further extended to explore regular rotating black holes and solitons \cite{dymnikova_orbits_2023}. Recently, a new concept of particle surfaces has been defined in Ref. \cite{2023arXiv230612888B}. These surfaces represent the collection of points swept by particles on spherical orbits. This notion can be generalized to other types of stationary spacetimes using the methods provided in our paper and can be used to delimit the regions of particles around black holes. These are just a few examples of potentially interesting subjects that can be investigated in future works, utilizing the numerical and analytical methods discussed in this paper. }

\section*{Acknowledgements}
MF and NC acknowledge Universidad de Santiago de Chile for financial support through the Proyecto POSTDOC-DICYT, C\'{o}digo 042331CM$\_$Postdoc. JRV is partially supported by the Centro de Astrof\'isica de Valpara\'iso (CAV).

\appendix

\section{The roots of the polynomial equation $\Delta(r)=0$}\label{app:A}

By employing the method outlined in Appendix B of Ref. \cite{fathi_spherical_2023}, the solutions to the suppressed quartic equation \eqref{eq:Delta=0} are derived as follows:
\begin{eqnarray}
&& r_1 = \frac{1}{\Lambda}\left[\tilde{\mathcal{A}}+\sqrt{\tilde{\mathcal{A}}^2-\tilde{\mathcal{B}}}\,\right]+\frac{3\gamma}{4\Lambda},\label{eq:r1}\\
&& r_2 = \frac{1}{\Lambda}\left[\tilde{\mathcal{A}}-\sqrt{\tilde{\mathcal{A}}^2-\tilde{\mathcal{B}}}\,\right]+\frac{3\gamma}{4\Lambda},\label{eq:r2}\\
&& r_3 = \frac{1}{\Lambda}\left[-\tilde{\mathcal{A}}+\sqrt{\tilde{\mathcal{A}}^2-\tilde{\mathcal{C}}}\,\right]+\frac{3\gamma}{4\Lambda},\label{eq:r3}\\
&& r_4 = \frac{1}{\Lambda}\left[-\tilde{\mathcal{A}}-\sqrt{\tilde{\mathcal{A}}^2-\tilde{\mathcal{C}}}\,\right]+\frac{3\gamma}{4\Lambda},\label{eq:r4}
\end{eqnarray}
in which 
\begin{subequations}
\begin{align}
    & \tilde{\mathcal{A}} = \sqrt{\tilde{\mathcal{U}}-\frac{\mathcal{A}}{6}},\\
    & \tilde{\mathcal{B}} = 2\tilde{\mathcal{A}}^2+\frac{\mathcal{A}}{2}+\frac{\mathcal{B}}{4\tilde{\mathcal{A}}},\\
    & \tilde{\mathcal{C}} = 2\tilde{\mathcal{A}}^2+\frac{\mathcal{A}}{2}-\frac{\mathcal{B}}{4\tilde{\mathcal{A}}},
\end{align}
\label{eq:tAtBtC}
\end{subequations}
where 
\begin{subequations}
\begin{align}
    & \mathcal{A} = -\left(\frac{27 \gamma^2}{8 \Lambda^3}+\frac{3}{\Lambda^2}\right)\Lambda^3,\\
    & \mathcal{B} = \left(\frac{6M}{\Lambda}-\frac{27\gamma^3}{8\Lambda^3}-\frac{9\gamma}{2\Lambda^2}\right)\Lambda^3,\\
    & \mathcal{C} =\left(\frac{2M \gamma}{2\Lambda}-\frac{27\gamma^2}{16 \Lambda^2}-\frac{243\gamma^4}{256\Lambda^3}-3a^2\right)\Lambda^3,\label{eq:mathcalC}
\end{align}
\label{eq:ABC}
\end{subequations}
and
\begin{equation}
\tilde{\mathcal{U}} = \sqrt{\frac{g_2}{3}}\cosh\left(\frac{1}{3} \arccosh\left(3g_3\sqrt{\frac{3}{g_2^3}}\right)\right),
    \label{eq:tU}
\end{equation}
with
\begin{subequations}
\begin{align}
    & g_2 = \frac{\mathcal{A}^2}{48}+\frac{\mathcal{C}}{4},\\
    & g_3 = \frac{\mathcal{A}^3}{864}-\frac{\mathcal{A}\mathcal{C}}{24}+\frac{\mathcal{B}^2}{64}.
\end{align}
\label{eq:g2g3}
\end{subequations}

\bibliographystyle{ieeetr}
\bibliography{biblio_v1.bib}

\end{document}